\documentclass[lettersize, final, twoside, journal]{IEEEtran}
\usepackage{amssymb, amsmath, amsthm, amsfonts}
\usepackage{bm, color}
\usepackage{booktabs}
\usepackage{algorithm}
\usepackage{algorithmic}
\usepackage{setspace}
\usepackage{url}
\usepackage{graphicx}
\usepackage{longtable, threeparttable}
\usepackage{makecell}
\usepackage{multirow}
\usepackage{ragged2e}
\usepackage{fancyhdr}
\usepackage{bbding}
\usepackage{cite}
\usepackage{dsfont}
\usepackage{datetime}
\usepackage{float}
\usepackage{siunitx}

\newtheorem{proposition}{Proposition}

\allowdisplaybreaks[4]

\usepackage{multicol}  
\usepackage{wrapfig}
\usepackage{float}
\usepackage[caption=false, font=footnotesize]{subfig}

\newcommand{\tabincell}[2]{\begin{tabular}{@{}#1@{}}#2\end{tabular}}

\begin{document}

\title{A CPFSK Transceiver with Hybrid CSS–DSSS Spreading for LPWAN PHY Communication}

\author{Wenkun~Wen,~\IEEEmembership{Member,~IEEE}, Ruiqi~Zhang, Peiran~Wu,~\IEEEmembership{Member,~IEEE}, Tierui Min, \\ and Minghua~Xia,~\IEEEmembership{Senior Member,~IEEE}

\thanks{Manuscript received 13 July 2025; revised 25 July 2025; accepted 29 July 2025. \textit{(Corresponding author: Minghua Xia.)}

Wenkun Wen and Tierui Min are with Techphant Technologies Co. Ltd., Guangzhou 510310, China (e-mail: wenwenkun@techphant.net, mintierui@techphant.net)

Ruiqi Zhang, Peiran Wu, and Minghua Xia are with the School of Electronics and Information Technology, Sun Yat-sen University, Guangzhou 510006, China (e-mail: zhangrq7@mail2.sysu.edu.cn, wupr3@mail.sysu.edu.cn, xiamingh@mail.sysu.edu.cn).

Color versions of one or more of the figures in this article are available online at https://ieeexplore.ieee.org.
	
Digital Object Identifier 

Copyright (c) 2025 IEEE. Personal use of this material is permitted. However, permission to use this material for any other purposes must be obtained from the IEEE by sending a request to pubs-permissions@ieee.org.
}
}

\markboth{IEEE Internet of Things Journal} {Wen \MakeLowercase{\textit{et al.}}: A CPFSK Transceiver with Hybrid CSS–DSSS Spreading for LPWAN PHY Communication}

\maketitle

\IEEEpubid{\begin{minipage}{\textwidth} \ \\[12pt] \centering 2327-4662 \copyright\ 2025 IEEE. Personal use is permitted, but republication/redistribution requires IEEE permission. \\
See \url{https://www.ieee.org/publications/rights/index.html} for more information.\end{minipage}}

 \IEEEpubidadjcol

\begin{abstract}
Traditional low-power wide-area network (LPWAN) transceivers typically compromise data rates to achieve deep coverage. This paper presents a novel transceiver that achieves high receiver sensitivity and low computational complexity. At the transmitter, we replace the conventional direct sequence spread spectrum (DSSS) preamble with a chirp spread spectrum (CSS) preamble, consisting of a pair of down-chirp and up-chirp signals that are conjugate to each other, simplifying packet synchronization. For enhanced coverage, the payload incorporates continuous phase frequency shift keying (CPFSK) to maintain a constant envelope and phase continuity, in conjunction with DSSS to achieve a high spreading gain. At the receiver, we develop a double-peak detection method to improve synchronization and a non-coherent joint despreading and demodulation scheme that increases receiver sensitivity while maintaining simplicity in implementation. Furthermore, we optimize the preamble detection threshold and spreading sequences for maximum non-coherent receiver performance. The software-defined radio (SDR) prototype, developed using GNU Radio and USRP, along with operational snapshots, showcases its practical engineering applications. Extensive Monte Carlo simulations and field-test trials demonstrate that our transceiver outperforms traditional ones in terms of receiver sensitivity, while also being low in complexity and cost-effective for LPWAN requirements.
\end{abstract}


\section{Introduction}
\label{Introduction}
\IEEEPARstart{T}{O} implement Internet of Things (IoT) networks, massive machine-type communications (mMTC) are committed to information exchange among enormous end devices characterized by short packets, discontinuous data transmission, low-power consumption, high-density deployment, and deep coverage \cite{mMTC}. Accordingly, several low-power wide-area network (LPWAN) technologies have been developed to connect massive devices distributed over large areas at a low cost and with low power consumption. Among them, two typical commercial solutions are NB-IoT, operating in licensed frequency bands, and LoRa, working in unlicensed ones. The former utilizes orthogonal frequency division multiplexing (OFDM) as its physical layer (PHY) modulation scheme, while the latter adopts chirp spread spectrum (CSS). Although performing well in deep coverage thanks to advanced receiver design, these two LPWAN technologies struggle to achieve relatively low-cost end-device implementation \cite{9778216}. Specifically, NB-IoT devices rely on cellular modem modules that are significantly more complex than typical microcontrollers, as they must comply with 3GPP standards and support a full cellular protocol stack—factors that contribute to increased hardware costs. In contrast to NB-IoT, which is fully standardized and provides open access to detailed physical layer (PHY) specifications \cite{11017626}, LoRa employs a proprietary PHY layer patented by Semtech Corporation, making it more challenging to conduct in-depth research into its physical layer performance \cite{10609524}. Furthermore, to ensure full ecosystem interoperability, LoRaWAN end devices must undergo certification by the LoRa Alliance, which, coupled with the potential need for privately deployed gateways, can result in elevated network infrastructure costs.

\subsection{Related Works and Motivation}
Compared to the OFDM inherent in NB-IoT, which has a high peak-to-average power ratio (PAPR), narrowband continuous phase frequency shift keying (CPFSK), such as minimum shift keying (MSK), is more suitable for low-cost IoT end devices \cite{10683565}. On the one hand, CPFSK is a constant envelope modulation (CEM) with both constant amplitude and phase continuity, allowing the transceiver power amplifiers (PAs) to operate at or close to the saturation level \cite{9990567}. On the other hand, the CPFSK receiver can be designed with low complexity because the received symbol can be demodulated non-coherently \cite{1092333}, which eliminates the channel estimation phase and renders it insensitive to residual carrier frequency offsets (CFOs) and sampling time offsets (STOs) \cite{NPRACH, 10844041}. 

\IEEEpubidadjcol
 
To meet the deep coverage requirement of mMTC, the direct sequence spread spectrum (DSSS) technology, which benefits coverage extension and data rate adaptation through its spreading gain \cite{LPWAN2}, can be integrated with CPFSK for a low-complexity LPWAN PHY design, abbreviated as DSSS-CPFSK PHY. Compared to the CSS PHY inherent in LoRa, the DSSS-CPFSK PHY meets the more comprehensive range requirement of data rate adaptation and coverage extension, thanks to the more flexible selection of spreading factors and coding rate/type. Unlike LoRa, which involves two additional transceiver schemes \cite{sx1272}, the DSSS-CPFSK PHY-based transceiver can integrate only one modulation scheme, resulting in lower complexity and cost at the end device. In addition, unlike CSS, whose modulation order takes the values $2^{\rm SF}$, with the limited spreading factor $\rm SF \in \{7, \cdots, 12\}$, CPFSK has a much lower modulation order and uses fewer distinct frequency states to represent data symbols, allowing for the use of simpler frequency discriminators or non-coherent detectors that do not require complex phase tracking, thereby reducing the cost of its end-device receiver.

LPWAN communications are typically short, packet-based, discontinuous, and asynchronous, which necessitates the efficient design of preamble detection to identify the start of packets (SOP). The DSSS-based IoT PHY adopts DSSS sequences with good autocorrelation as its preamble \cite{IEEE802154}. However, the large CFOs caused by inaccurate crystal oscillators in the end devices ultimately destroy the autocorrelation, resulting in significant difficulty in preamble detection. Autocorrelation/differential correlation is a low-complexity and CFO-insensitive preamble detection scheme \cite{xu2017design, DSSS_Diff}. It first multiplies received samples by local reference and then calculates the autocorrelation of the processed sequence, with the peak indicating the SOP. However, the multiplication of received samples will amplify the noise greatly, which makes this scheme only work well at high SNR. Double correlation is a high-performance detection method of the DSSS preamble, which calculates the autocorrelation of the multiplied samples with different lags and combines their magnitudes to suppress noise \cite{DC, DC1, MSK_Sync}. Although robust to CFOs and works well at low SNR, its computational complexity is too high to be practical for low-cost IoT terminals \cite{8657965}. 

Compared to the DSSS-CPFSK signal, chirp signals are much more CFO-tolerant while still maintaining good correlation properties, making them more suitable as a preamble in low-cost IoT PHY. The complete inverse engineering of the LoRa PHY \cite{EPFL} revealed that its preamble consists of $8$ repetitive upchirps by default and $2.25$ downchirps, which accommodates its extremely high modulation order and low sensitivity levels. A more extended sequence of preamble chirps allows for precise CFO and STO estimation, but it also increases the sequence transmission energy and the processing latency. In real-world applications, the fractional parts of CFO and STO can mostly lead to a symbol being incorrectly interpreted as one of its adjacent symbols \cite{Seller2016, 9154273, 9000820, 9501038}. This issue is referred to as the `±1 demodulation error', which can be effectively corrected using Gray mapping \cite{9154273, 9148806}. Moreover, it is demonstrated in \cite{9501038} that the fractional CFO smaller than $0.1$ has almost no impact on the LoRa receiver. Consequently, fewer preamble chirps can be used in scenarios involving low modulation orders and reduced estimation accuracy for the fractional parts of CFO and STO. Additionally, a shorter preamble facilitates quicker synchronization, decreases energy overhead, and reduces processing latency.

The combined use of upchirps and downchirps in the preamble allows for separate estimation of the CFO and the STO, because the STO has an opposite effect on upchirps and downchirps \cite{9000820, 9501038}. Thus, the most effective preamble consists of two base chirps conjugate to each other. In this context, the pioneering work \cite{Chirp_Sync1} demonstrates that a linear chirp signal followed by its complex conjugate is an efficient reference preamble for frame detection and synchronization. The estimation of time offsets is further improved in \cite{Chirp_Sync3}. The idea of using two chirps has been expanded to tackle scenarios involving frequency-selective fading \cite{4783007}, where the training sequence comprises one upchirp and two downchirps. The extra downchirp serves to fine-tune the fractional frequency offset. Moreover, a more compact dual-chirp, comprising a linear upchirp transmitted simultaneously with its complex conjugate downchirp, has proven effective in frequency-selective fading channel scenarios \cite{6775034, 9083764, 10498090}.

This work aims to develop a low-cost and low-complexity transceiver while ensuring comparable performance. Common approaches to improving spectral efficiency and/or energy efficiency of LoRa systems include double-mode CSS \cite{9828505}, dual-mode time-domain multiplexed CSS \cite{10183362}, and differential multi-mode CSS \cite{10296020}. For readers interested in further information, a recent survey is available in \cite{10391276}.

\subsection{Contributions}
This paper designs an end-to-end transceiver for LPWAN applications based on narrowband CEM, which offers higher receiver sensitivity compared to the state-of-the-art DSSS-CPFSK PHY \cite{xu2017design, FSK_DSSS} while maintaining relatively low synchronization complexity. In particular, we utilize different CEM schemes in the preamble and payload to reduce receiver complexity. Specifically, CSS with two base chirps, conjugate to each other, is adopted as the preamble, while DSSS-CPFSK is applied to the payload. In brief, four significant contributions include:
\begin{itemize}
	\item Based on a minimal preamble consisting of a pair of conjugate chirps, an efficient preamble detection and coarse synchronization scheme is developed, which jointly estimates the SOP and coarse CFO through the cross-correlation peak-value positions of down- and up-chirps. This scheme is viable as it allows for a constant false alarm rate (CFAR) without knowledge of the actual received SNR.
	\item A non-coherent joint despreading and demodulation scheme is devised for the DSSS-CPFSK payload, which performs symbol-level non-coherent detection to achieve higher spreading gain than existing schemes \cite{xu2017design, FSK_DSSS}. It has lower computational complexity than coherent receivers and is robust to phase rotation. 
	\item To achieve the target false alarm rate, the optimal preamble detection threshold is explicitly derived (see Proposition~\ref{Proposition-1}). Additionally, to ensure optimal payload demodulation performance, the criterion for selecting the spreading sequences is established (see Proposition~\ref{Proposition-2}).
	\item To evaluate the performance of our transceiver design in practice, we develop a software-defined radio (SDR) prototype based on the GNU Radio toolkit and USRP. Extensive Monte-Carlo simulation experiments and field-test trials demonstrate the effectiveness of our transceiver design and prototype development. 
\end{itemize}

The rest of this paper is organized as follows. Section~\ref{SystemModel} describes the transceiver architecture, including the block diagram and signal models. Section~\ref{Section-Chirp_analysis} reviews the correlation properties of chirp signals and highlights their capability of CFO-tolerance. Section~\ref{ReceiverDesign} elaborates on a low-complexity receiver, including synchronization and demodulation steps. Section~\ref{PrototypeDevelopment} develops a prototype using the GNU Radio SDR toolkit and USRP, with operational snapshots illustrating. Section~\ref{SimulationResults} presents and discusses Monte-Carlo simulation results in MATLAB, compared to the prototype field-test results. Finally, Section~\ref{ConcludingRemarks} concludes the paper.

\begin{figure*}[t]
	\centering
	\centerline{\includegraphics[width = 0.95\textwidth]{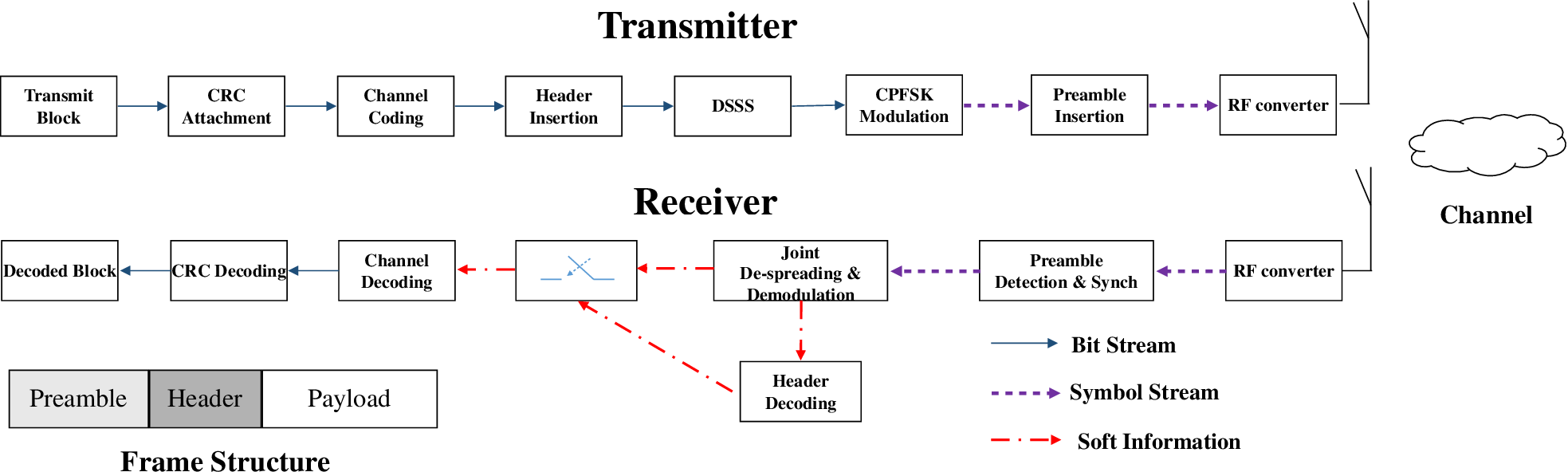}}
	\vspace{-5pt}
	\caption{The PHY block diagram of the designed transceiver.}
	\label{PHY Transceiver}
	\vspace{-10pt}
\end{figure*}

\section{System Model} 
\label{SystemModel}
This section starts with the frame structure and corresponding signal models, followed by the complete transceiver chain.

\subsection{Frame Structure and Signal Models}
The frame structure of our PHY design consists of three parts: preamble, header, and payload, as shown in the left lower corner of Figure~\ref{PHY Transceiver}. Among them, the preamble is used for packet presence detection and synchronization. The header contains some important PHY parameters, including the packet length, code rate, presence of a cyclic redundancy check (CRC), checksum, and modulation type. Finally, the payload bears the data to be transmitted. Next, we describe the signal model of each part.

\subsubsection{\underline{Preamble}} \label{SectionII-A}
We adopt digital chirp signals as the preamble due to their robustness against high CFOs, which will be further discussed in Section~\ref{Section-Chirp_analysis}. Using an approach similar to \cite[Eq. (13)]{8723130}, a base down-chirp symbol with swept-bandwidth $B$ and spreading factor $\rm SF$, can be equivalently expressed in discrete-time baseband model as
\begin{align} \label{down-chirp}
s_{\rm d}[n] 
	= \exp\left(-j 2\pi \left(\frac{ n^{2} }{2N K^{2}} - \frac{n}{2K}\right)\right), 
\end{align}
for all $n = 0, 1, \cdots, NK-1$, where $N = 2^{\rm SF}$ denotes the chip number in one chirp symbol, $K$ is the oversampling factor for one chip, and the offset $-n/(2K)$ is used to constrain the frequency sweep range in $\left[-B/2, B/2\right]$ \cite[Eq. (5)]{8723130}. The complex conjugation of the base down-chirp $s_{\rm d}[n]$, denoted $s_{\rm u}[n] = s_{\rm d}^{*}[n]$, is called base up-chirp with linearly increasing swept-frequency. 

Our preamble consists of a linear down-chirp followed by an up-chirp with $2N$ chips in total, given by
\begin{equation} \label{preamble}
	s_{\rm pre}[n] = \left\{ \hspace{-5pt}
		\begin{array}{rl}	
			s_{\rm d}[n], & \hspace{-3pt} n = 0,1, \cdots, NK-1; \\
			s_{\rm u}[n- NK], & \hspace{-3pt} n = NK, \cdots, 2NK-1.
		\end{array} \right.
\end{equation}
Based on this concise preamble structure, packet detection and synchronization can be devised efficiently, as elaborated in Section~\ref{ReceiverDesign} shortly.

\subsubsection{\underline{Header/Payload}} \label{SectionII-B}
The $L$-bit transmit data after channel coding is first spread by DSSS codes with ${\rm SF}_{\rm p}$-chip length and then is modulated by a binary CPFSK, given by \cite{DC}
\begin{align} \label{CPFSK_Phase}
	s_{\rm pay}[n] 
		= \exp\Big(j2 \pi h_{\rm m} \sum_{i = 0}^{L {\rm SF}_{\rm p}-1}\alpha[i] \, q[n-iK]\Big),
\end{align}
where $\alpha[i] \in \left\{+1, -1\right\}$ is the $i^{\rm th}$ binary code of the processed sequence $\bm{\alpha}$ with the length of $L {\rm SF}_{\rm p}$ chips; $q[n]$ is the phase response of CPFSK and is represented by the cumulation of the rectangular phase-shaping pulse with one-chip duration; $h_{\rm m} = \Delta f_{\rm dev} / B$ is the modulation index of CPFSK, where $\Delta f_{\rm dev}$ denotes the frequency deviation between two frequency points. As larger $h_{\rm m}$ means larger frequency deviation and larger occupied bandwidth, the oversampling factor $K$ must be sufficiently large to satisfy the minimum Nyquist rate requirement, i.e., $K \geq \lceil 2 h_{\rm m} \rceil$. For implementation compatibility, the preamble, header, and payload adopt the same chip and sampling rates at both the transmitter and receiver, i.e., their oversampling factors are identical. 

Combining \eqref{preamble} and \eqref{CPFSK_Phase}, the transmit signals can be explicitly expressed as
\begin{equation} \label{Eq-Tx}
\small
	s[n] = \left\{\hspace{-5pt}
		\begin{array}{rl}		
			s_{\rm pre}[n] , & \hspace{-0.5em} n = 0,1, \cdots, 2NK-1; \\
			s_{\rm pay}[n- 2NK], & \hspace{-0.5em} n = 2NK, \cdots, (2N+ L {\rm SF}_{\rm p})K-1.
		\end{array} \right. \nonumber
\end{equation}

As the MSK signal has higher spectral efficiency than orthogonal FSK, which allows a decrease in the deployment intervals of adjacent channels and/or a reduction of the thermal noise at the receiver, we adopt MSK in our design as the primary modulation scheme, combined with DSSS technology and channel coding to improve the receiver sensitivity. It is noteworthy that, compared to LoRa, which employs a $2^{\text{SF}}$-ary modulation scheme, the binary modulation of MSK results in lower spectral efficiency. To achieve higher spectral efficiency while preserving the continuous phase and constant envelope properties, $M$-ary CPFSK modulation can be employed, though at the cost of increased demodulation complexity and potential sensitivity to channel impairments.

\subsection{Transceiver Chain}
Figure~\ref{PHY Transceiver} depicts the block diagram of our transceiver. At the transmitter shown in its upper panel, CRC bits are first calculated and appended to the end of each raw bit package for error checking. Then, the convolutional coding, scrambling, and interleaving operations are applied as channel coding for extra payload protection. Next, the coded header is inserted before the coded payload, and then the processed sequence is spread by the DSSS code with spreading factor ${\rm SF}_{\rm p}$. Apart from the payload symbols, the complex preamble samples for frame detection and synchronization are also generated and concatenated before transmission. Finally, the baseband signals are modulated into RF signals and radiated through a Tx antenna.

At the receiver shown in the lower panel of Figure~\ref{PHY Transceiver}, the received RF signals are first converted to the baseband ones. Suppose the wireless channel is frequency-flat and time-invariant in an LPWAN application with a low data rate and mobility, the received baseband samples can be expressed as
\begin{equation} \label{rx_signal}
	r[n] = hs[n-\tau]e^{ j 2 \pi \Delta f \left(n-\tau\right)} + w[n], \nonumber
\end{equation}
for all $n = 0,1,\cdots, (2N+ L {\rm SF}_{\rm p})K-1$, where $h = |h|e^{j\theta_{h}}$ denotes the channel fading coefficient with magnitude $|h|$ and phase $\theta_{h}$\footnote{In LPWAN applications, CSS symbols typically maintain a narrow bandwidth of $500$ kHz or less \cite{10183362}. In our prototype development, we use a USRP operating at a central frequency of $f_c = 470$ MHz with a swept bandwidth of $B = 76.8$ kHz. Given the narrow bandwidth and stationary terminals, we assume a frequency-flat fading channel in the paper.}; $\Delta{f} \triangleq \Delta_{\rm CFO}/f_{s}$ denotes the normalized CFO w.r.t. the sampling frequency $f_{s} = K B$, which has an integer part $\Delta f_{\rm I}$ and a fractional one $\Delta f_{\rm F}$ of the resolution of coarse CFO estimation (i.e., $\Delta{f} = \Delta{f_{\rm I}} + \Delta{f_{\rm F}}$); $\tau$ is the STO normalized w.r.t. sampling duration at the receiver, consisting of an integer part $\mu$ and a fractional one $\epsilon$ (i.e., $\tau = \mu + \epsilon$);  $w[n] \sim \mathcal{CN}(0, \sigma^{2})$ refers to the circularly symmetric AWGN, with $\sigma^{2}$ being the noise variance.
 
At the receiver, synchronization is critical for packet detection and signal recovery. In our design, with the aid of chirp signals in the preamble, synchronization can be achieved efficiently using a simple sliding correlation method.  Then, we perform header demodulation to recover the original header information. If the recovered header passes its CRC check, its inherent information is applied to assist the subsequent demodulation and decoding; otherwise, the header will be dropped, and the receiver will terminate demodulation of the payload. The header and the payload adopt the non-coherent joint despreading and demodulation scheme, which will be designed shortly. Finally, the de-interleaving, descrambling, and Viterbi soft-decoding are performed sequentially to achieve further gains in channel coding. The final CRC decoding indicates whether the package is received successfully or not.

\section{Correlation and CFO-Tolerance of Down- and Up-Chirp Signals}  
\label{Section-Chirp_analysis}
This section provides a brief review of the correlation properties of down-chirp and up-chirp signals, demonstrating their superior capability for CFO tolerance compared to conventional DSSS-CPFSK signals.

By \eqref{down-chirp}, the autocorrelation magnitudes $\left| \rho_{\rm dd}[m] \right|$ of a linear down-chirp $s_{\rm d}[n]$ can be readily computed as
\begin{align} \label{autocorrelation}
	\left| \rho_{\rm dd}[m] \right| 
	= \frac{1}{NK}\left|\frac{\sin \left(\frac{\pi m}{NK^{2}} \left( NK - |m|\right) \right) }{\sin\left(\frac{\pi m}{NK^{2}} \right)} \right|,
\end{align}
for all timing lags $ m = 0,\pm 1,\cdots,\pm(NK-1)$. As \eqref{autocorrelation} gives the absolute values of a discrete Sinc function centered at $m = 0$, it takes the maximum value at $m = 0$ while approximating zero otherwise, indicating the excellent autocorrelation property of chirp signals. Likewise, the cross-correlation magnitudes $\left|\rho_{\rm du}[m]\right|$ between a linear down-chirp $s_{\rm d}[n]$ and its complex conjugate can be computed and expressed as  
\begin{align}	\label{cross-correlation}
\left|\rho_{\rm du}[m]\right|  
	 = \frac{1}{NK} \left|\sum_{n = 0}^{NK - \left|m\right|-1} \hspace{-1.5em}
\exp\left(j2\pi\left(\frac{n^2+|m|n}{NK^2} - \frac{n}{K}\right)\right)\right|,
\end{align}
for all $ m = 0,\pm 1,\cdots,\pm(NK-1)$, which approaches zero for large $N$ \cite{8723130}. Therefore, the down- and up-chirp signals have an excellent cross-correlation property. 

Compared to the DSSS-based preamble, the chirp signal is much more CFO-tolerant and is ideal for the preamble. Specifically, in the case of the CFO, say, $\Delta f$, the cross-correlation amplitudes between a linear down-chirp and its copy can be readily expressed as
\begin{align} \label{correlation}
	\lefteqn{\left| \rho_{\rm dd}[m, \Delta f] \right|} \nonumber \\
		& = \frac{1}{NK}\left|\sum_{n} s^{\ast}_{\rm d}[n]s_{d}[n+m]e^{j 2\pi \Delta f (n+m)} \right| \notag \\
	& = \frac{1}{NK}\left| \frac{ \sin \left(\frac{\pi m'}{NK^{2}} \left(N K \left(1 - \left|K \Delta f\right| \right) - |m'| \right)\right)} {\sin \left( \frac{\pi m'}{NK^{2}} \right)} \right|,
\end{align}
where $m' \triangleq m - NK^{2}\Delta f$, with $ m = 0,\pm 1,\cdots,\pm(NK-1)$. Clearly, \eqref{correlation} resembles \eqref{autocorrelation}, but with the peak-value center shifting from $0$ to $\lceil NK^{2}\Delta f\rceil$. This shift is commonly referred to as the timing-frequency coupling feature. When using the Nyquist sampling interval (i.e., $T_{s} = 1/B$), the actual peak falls between $\lceil NK^{2}\Delta f\rceil$ and its neighboring values, which yields peak leakage. As a result, applying an oversampling factor $K > 1$ is advantageous in practice for reducing the effects of peak leakage. Similarly, the magnitudes of the cross-correlation between a linear up-chirp signal $s_{\rm u}[n]$ and its signal with added CFO can also be computed by \eqref{correlation}. In this case, the peak-value center shifts from $\lceil NK^{2}\Delta f\rceil$ to its inverse value.

\begin{figure}[!t] 
	\centerline{\includegraphics[width=0.45\textwidth]{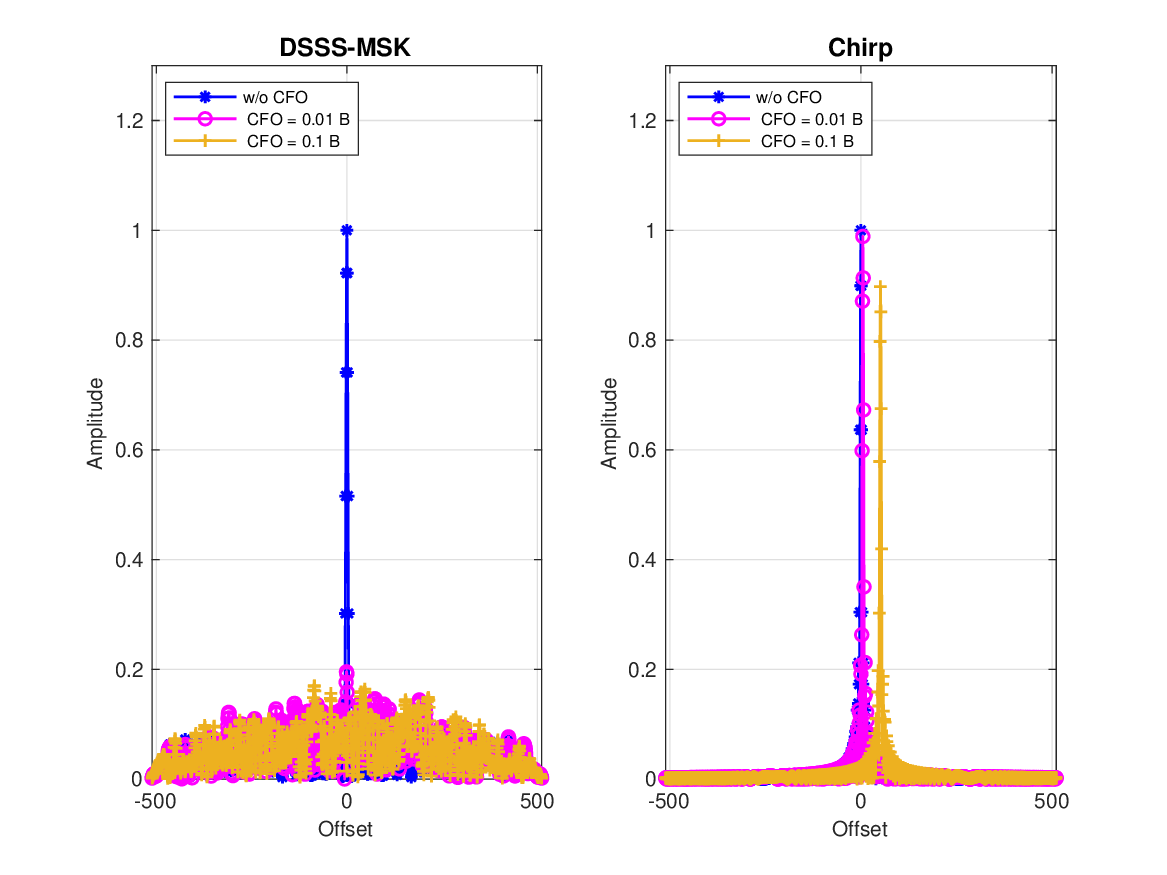}}
	\vspace{-10pt}
	\caption{CFO-tolerance capability of DSSS-MSK and Chirp signals.}
	\label{cross_corr}
\end{figure}

For illustration, Figure~\ref{cross_corr} depicts the cross-correlation amplitudes of the DSSS-MSK and chirp signals, with a chip length $N = 128$ and an oversampling factor $K = 4$. The left subfigure indicates that even a small CFO, normalized to $1\%$ of the bandwidth, significantly degrades the cross-correlation peak amplitude for the DSSS-MSK signal. In contrast, the right subfigure illustrates that even a large CFO, normalized to $10\%$ of the bandwidth, has a negligible impact on the amplitude of the cross-correlation peak of the chirp signal. This tolerance to the CFO is crucial for IoT applications, especially for end devices equipped with low-cost and imprecise crystal oscillators.

Based on \eqref{correlation}, the maximum value of $\rho_{\rm dd}[m, \Delta f]$ can be readily computed as
\begin{equation} \label{Eq-MaxXcorr}
	\rho_{\max}(K, \Delta{f}) = \frac{1}{NK}\left| \frac{ \sin\left(\frac{\pi}{NK^{2}} \Delta{\epsilon} (NK (1 - \left|\Delta{f}\right|) - |\Delta \epsilon|)\right)} {\sin \left( \frac{\pi}{NK^{2}} \Delta{\epsilon} \right)} \right|,
\end{equation}
where $\Delta{\epsilon} \triangleq \lceil NK\Delta{f} \rfloor - NK\Delta{f}$. 

\begin{figure}[!t]
	\centering
	\centerline{\includegraphics[width = 0.55\textwidth]{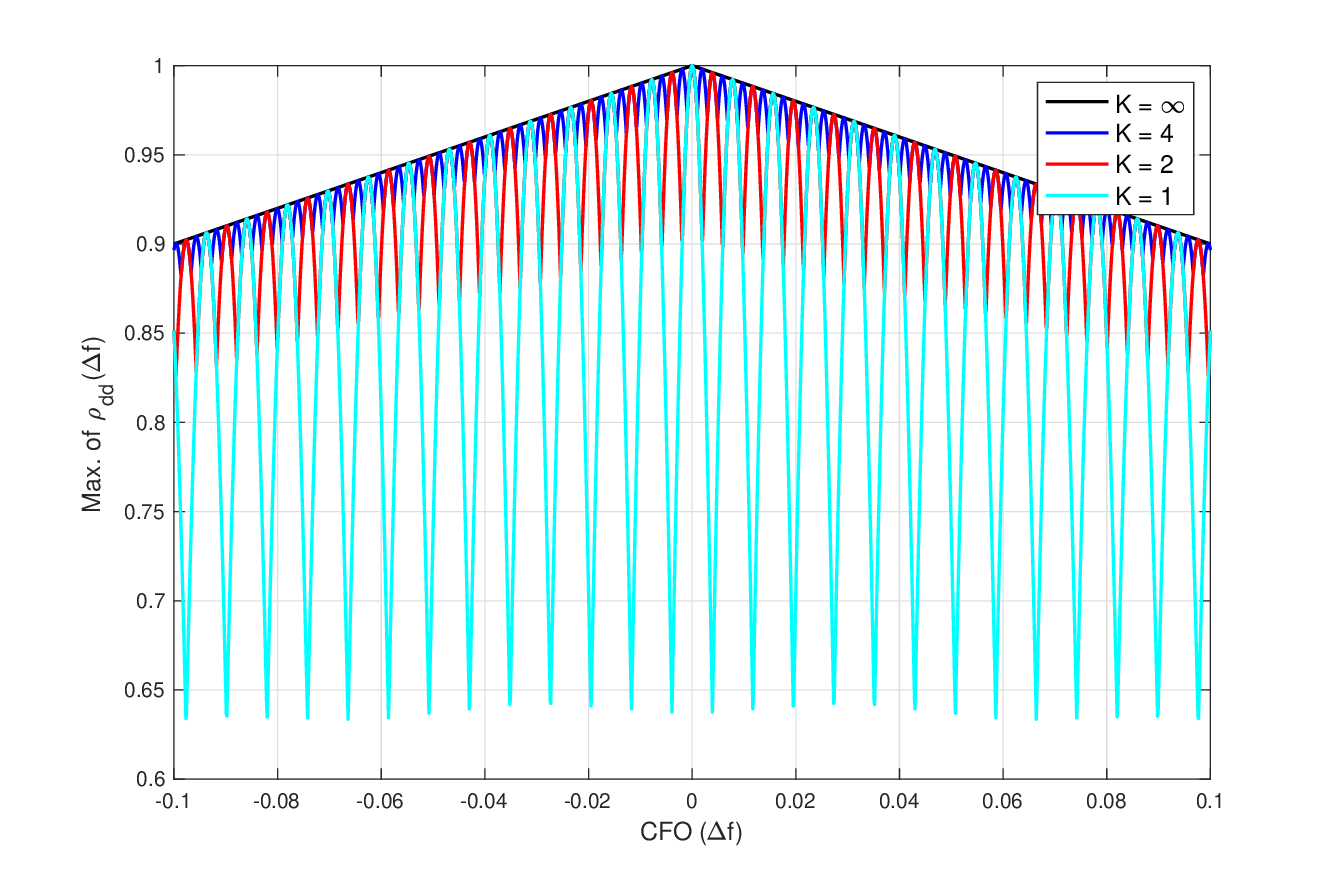}}
	\vspace{-10pt}
	\caption{The maximum of the cross-correlation coefficient versus the normalized CFO.}
	\label{Fig_MaxXcorr}
\end{figure}

Fig.~\ref{Fig_MaxXcorr} illustrates the values of $\rho_{\max}(K, \Delta{f})$ as defined in \eqref{Eq-MaxXcorr} versus the normalized CFO for varying oversampling factors $K$. The case of $K = \infty$ represents the ideal correlation peak achievable in the analog domain. As shown, increasing $K$ reduces the fluctuation in the correlation peak, thereby enhancing CFO resilience.

\begin{figure}[!t]
	\centering
	\centerline{\includegraphics[width = 0.45\textwidth]{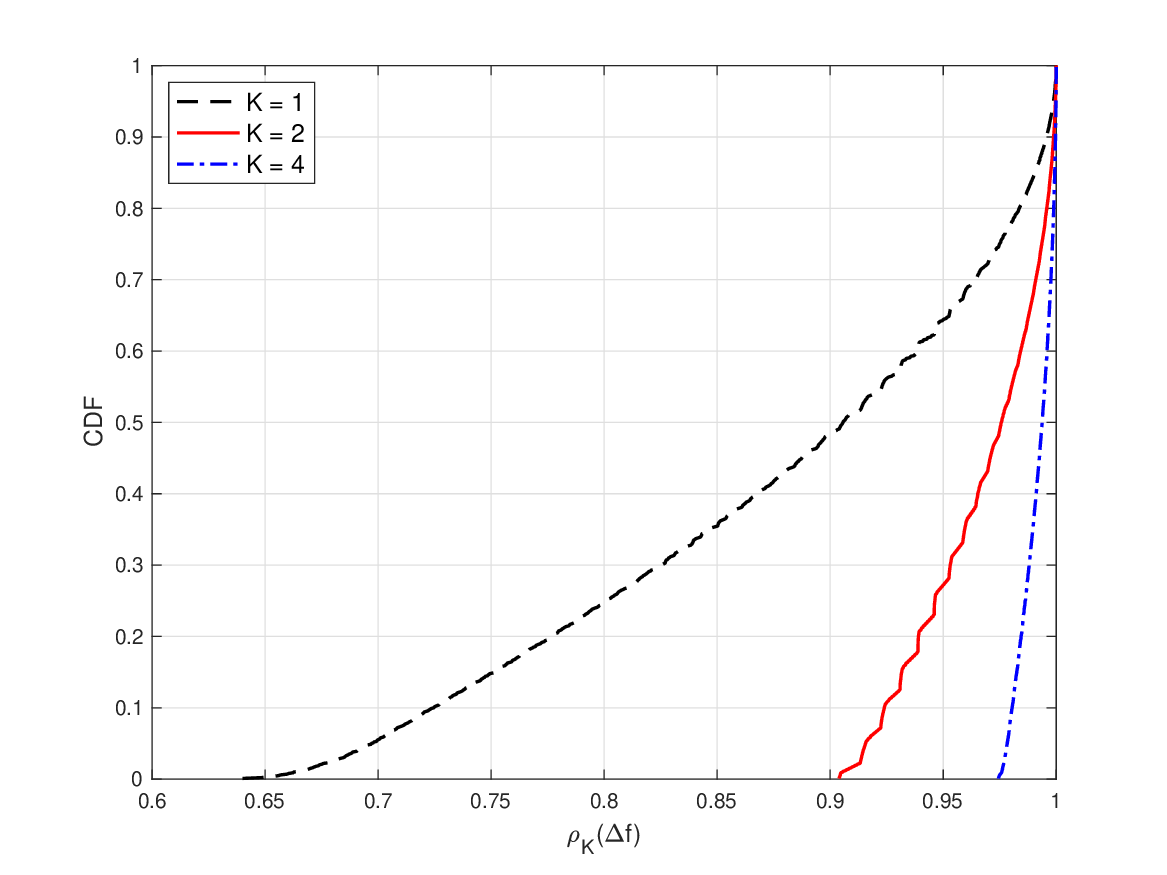}}
	\vspace{-10pt}
	\caption{CDF of $\rho_{K}(\Delta{f})$ defined in \eqref{Eq-XCorrRatio}.}
	\label{Fig_CDF}
\end{figure}

To quantify this improvement, we define the normalized correlation ratio:
\begin{equation}  \label{Eq-XCorrRatio}
	\rho_{K}(\Delta{f}) = \frac{\rho_{\max}(K, \Delta{f})}{\rho_{\max}(\infty, \Delta{f})}.
\end{equation}
Fig.~\ref{Fig_CDF} plots the cumulative distribution function (CDF) of $\rho_{K}(\Delta{f})$. It is observed that the minimum values (i.e., the smallest non-zero values where the CDF starts increasing) are approximately $0.65$, $0.9$, and $0.975$ when the oversampling factor $K$ increases from $1$ to $2$ and $4$, respectively. This implies that, with probability one, the values of $\rho_{K}(\Delta{f})$ exceed $0.65$, $0.9$, and $0.975$ in the respective cases. 

Clearly, increasing $K$ enhances the reliability of preamble detection by mitigating the impact of CFO. However, this improvement comes with an additional computational burden. 
To balance performance and complexity, we adopt $K = 2$, which offers a robust performance improvement with manageable overhead. For applications requiring even more resilient synchronization (e.g., under high mobility or extreme CFO), larger values like $K \ge 4$ are commonly used, as seen in practical LoRa implementations.

\section{Low-Complexity Receiver Design} \label{ReceiverDesign}
This section elaborates on three key designs at the receiver, including preamble detection and coarse synchronization, fine synchronization, and payload demodulation. Also, the Cramer-Rao lower bound (CRLB) is derived to evaluate the efficiency of CFO estimation. Next, we start with a double-peak approach for preamble detection and coarse synchronization.

\subsection{Double-Peak Approach for Preamble Detection and  Coarse Synchronization}
\label{Preamble_Detection}

At the receiver, the baseband signals after down-conversion are first oversampled with an oversampling factor $K > 1$, to allow the correction of fractional STO $\epsilon$ with higher precision $T_{\rm c}/(RK)$, where $T_{\rm c} = 1/B$ denotes one chip interval and $R$ refers to an upsampling factor. For each synchronization and demodulation step, however, the oversampled signals are decimated by a factor of $R$ to reduce their computational complexity. Consequently, synchronization and demodulation are performed at a sampling rate $f_s = KB$.

\subsubsection{\underline{Preamble Detection}}
The local reference signals of the down- and up-chirp in the preamble are applied for sliding cross-correlation separately to detect the peak-value positions of the down-chirp $\hat{\tau}_{\rm d}$ and the subsequent up-chirp $\hat{\tau}_{\rm u}$, respectively. By \eqref{correlation} and the accompanying analysis, when accounting for CFO, the real peak-value positions can be computed as
\begin{align}
	{\tau}_{\rm d} &= \mu - \lceil NK^2 \Delta f \rfloor, \label{timing_u1} \\
	{\tau}_{\rm u} &= \mu + NK + \lceil NK^2 \Delta f \rfloor. \label{timing_u2}
\end{align}
After locating these two peak-values, the estimates of SOP $\mu$ and coarse CFO $\Delta f_{\rm I}$ can be explicitly determined as
\begin{align}
	\hat{\mu} &= \frac{1}{2}\lceil\hat{\tau}_{\rm d} + \hat{\tau}_{\rm u} - NK\rfloor, \label{sto} \\
\Delta\hat{f}_{\rm I} &= \frac{\hat{\tau}_{\rm u}-\hat{\tau}_{\rm d}-NK} {2 NK^2}. \label{cfo}
\end{align}
Clearly, \eqref{cfo} implies that the resolution of the coarse CFO estimation is 
\begin{equation} \label{f_res}
	\Delta f_{\rm res} = \frac{1}{2 N K^2},
\end{equation}
which is inversely proportional to $K^2$. Therefore, the oversampling method is crucial to eliminate frequency aliasing. In reality, $K = 4$ is necessary for the LoRa application \cite{8723130, 9555814}, yielding a $6$ dB SNR gain.

Next, we utilize the concept of CFAR \cite{CFAR} to determine the detection threshold, as it is challenging to predetermine in the absence of channel state information during the initial phase of the receiver process. Specifically, the detection threshold is set proportional to the noise level of any position in the sliding window for down-chirp detection, which covers $W_{\rm d} = NK$ sample points or, equivalently, $N$ chips in time duration. Thus, the correlation at index $k$ can be computed as
\begin{equation} \label{corr}  
\left|\rho_{\rm d}\left[k \right]\right| = 
\frac{1}{N K}\left|\sum_{n=0}^{N K-1} r[n+k]s_{\rm d}^{*}[n]\right|, 
\end{equation}
for all $k = p_{1}, p_{1}+1, \cdots, p_{1}+W_{\rm d}-1$, with $p_{1}$ being the starting point of the sliding window. In practice, the fast Fourier transform (FFT)-based correlation approach can reduce computational complexity if the preamble consists of an extremely long chirp signal \cite{8903531}.  Consequently, an estimate of $\tau_{\rm d}$, denoted $\hat{\tau}_{\rm d}$, can be obtained by the maximum magnitude in the sliding window as
\begin{equation} \label{MLE}  
	\hat{\tau}_{\rm d} = \mathop{\arg\max}\limits_{ p_1 \leq k \leq p_1 + W_{\rm d}-1 } \left| \rho_{\rm d}\left[k\right] \right|.
\end{equation}  
Afterward, the detection test is performed by
\begin{equation} \label{PAR}
	\left | \rho_{\rm d}\left[\hat{\tau}_{\rm d}\right] \right| \mathop {\lessgtr }\limits _{\mathcal{H}_{1}}^{\mathcal{H}_{0}} \gamma \, \mathbb{E}\left[\left|\rho_{\rm d}[k]\right| \right], \, k \notin \{\hat{\tau}_{\rm d}-1, \hat{\tau}_{\rm d}, \hat{\tau}_{\rm d}+1\}
\end{equation} 
where $\mathcal{H}_{0}$ is the null hypothesis that the down-chirp is absent in this sliding window, whereas $\mathcal{H}_{1}$ is the alternative hypothesis that the down-chirp is present. By recalling \eqref{cross-correlation}, the cross-correlation between the received up-chirp and reference down-chirp is relatively small, such that the interference from the up-chirp can be ignored when detecting the down-chirp. On the other hand, by recalling the peak leakage effect, the correlation magnitudes of $k \in \{\hat{\tau}_{\rm d}-1, \hat{\tau}_{\rm d}, \hat{\tau}_{\rm d}+1\}$ are discarded for more tolerant noise estimation.  Finally, the scaling factor $\gamma$ in \eqref{PAR} is a constant that determines the miss and false alarm rates, and its value can be explicitly determined, as formalized in the following proposition.
\begin{proposition} \label{Proposition-1}
	The scaling factor $\gamma$ in \eqref{PAR} can be explicitly determined as
	\begin{equation}  \label{Eq-Prop-1}
		\gamma = \sqrt{-\frac{4}{\pi}\ln\left(1-\left(1-\sqrt{P_{\rm FA}}\right)^{\frac{1}{W}}\right)},
	\end{equation}
	where $P_{\rm FA}$ and $W$ denote the target false alarm rate and the sliding window width, respectively. 
\end{proposition}

\begin{IEEEproof}
	See Appendix~\ref{Appendix-A}.
\end{IEEEproof}

After the down-chirp $s_{\rm d}[n]$ is detected, the receiver turns to detect the subsequent up-chirp $s_{\rm u}[n]$. By recalling \eqref{timing_u1}-\eqref{timing_u2},
the range of correlation peak offset between $\tau_{\rm d}$ and $\tau_{\rm u}$ is 
\begin{equation} \label{koffset}
	\Delta{\tau} = NK \pm 2\lceil N K^{2}\left|\Delta f_{\max}\right| \rceil,
\end{equation}
where $\left|\Delta f_{\max}\right|$ represents the maximum frequency offset that the carrier frequency and the accuracy of the crystal oscillator can cause. Hence, the sliding window for $s_{\rm u}[n]$ is limited in the range centered at $\hat{\tau}_{\rm c} = \hat{\tau}_{\rm d} + NK$ with width $W_{\rm u}$, where $W_{\rm u}$ must be large enough to cover all of the possible peak-value positions of $\hat{\tau}_{\rm d}$, say, $W_{\rm u} > \lceil 4NK^2 \left|\Delta f_{\max}\right|\rceil$. Without loss of generality, we set $W_{\rm u} = W_{\rm d} = NK$, which implies that the maximum CFO estimation range normalized by $B$ is $0.25$. This range is practical as the maximum offset of $20$ ppm is typically expected for oscillators in commercial receivers \cite{9501038}.

On the other hand, the up-chirp signal $s_{\rm u}[n]$ can be detected in a similar way to detecting the down-chirp signal, as shown in \eqref{corr}-\eqref{PAR}. In particular, if \eqref{PAR} is not satisfied for $s_{\rm u}[n]$ detection, it means that the preamble is most likely to be detected with a false alarm so that the receiver has to restart to detect $s_{\rm d}[n]$. Otherwise, if $s_{\rm u}[n]$ is detected in $\hat{\tau}_{\rm u}$ with magnitude $\left|\rho_{\rm u}\left[\hat{\tau}_{\rm u}\right]\right|$, we finally need to compare the detected peak-values as per
\begin{equation} \label{Detection2}
	\left|\rho_{\rm u}\left[\hat{\tau}_{\rm u}\right]\right|
		\mathop {\lessgtr}\limits _{\mathcal{F}}^{\mathcal{T}}
			\eta \left|\rho_{\rm d}\left[\hat{\tau}_{\rm d}\right]\right|,
\end{equation}
where the scaling factor $\eta$ is a constant to decide whether $\left|\rho_{\rm d}\left[\hat{\tau}_{\rm d}\right]\right|$ has valid magnitude or not.  

Notably, since the SOP depends on the actual peak-value positions, an error in detecting $s_{\rm d}[n]$ yields false preamble detection. In this case, $\left\vert \rho_{\rm d}[\hat{\tau}_{\rm d}]\right\vert$ follows Rayleigh distribution whereas $\left\vert \rho_{\rm u}[\hat{\tau}_{\rm u}]\right\vert$ follows Rician distribution, which have significant magnitude difference. Accordingly, we compare two peak values and set the scaling factor $\eta$ to decide whether the first peak is detected, as shown in \eqref{Detection2}. If the magnitude of the first peak is below the threshold, the receiver will re-detect the down-chirp in the next sliding window. Clearly, a small value of $\eta$ reduces the error detection probability but makes the receiver re-detect the down-chirp more frequently, whereas a large one takes the opposite. Therefore, we set $\eta = 1.5$ in our Monte Carlo simulation experiments to balance a high false alarm rate and frequent re-detection.

\subsubsection{\underline{Coarse Synchronization}}
In \eqref{Detection2}, $\mathcal{T}$ implies that $\left|\rho_{\rm d}\left[\hat{\tau}_{\rm d}\right]\right|$ has nearly the same power as $\left|\rho_{\rm u}\left[\hat{\tau}_{\rm u}\right]\right|$ so that the up-chirp detection succeeds; $\mathcal{F}$ implies that the magnitude of $\left|\rho_{\rm d}\left[\hat{\tau}_{\rm d}\right]\right|$ is quite lower than that of $\left|\rho_{\rm u}\left[\hat{\tau}_{\rm u}\right]\right|$ so that the down-chirp $s_{\rm d}[n]$ may be detected with false alarm. If $\mathcal{T}$ is satisfied, which means the whole preamble is detected, the coarse synchronization can be determined by \eqref{sto}-\eqref{cfo}. Otherwise, the receiver will be triggered to re-detect $s_{\rm d}[n]$ in the next sliding window. The value of $\hat{\tau}_{\rm d}$ will be updated if the newly detected peak value $\rho_{\rm d}[\hat{\tau}^{\prime}_{\rm d}]$ is larger than the previously detected $\rho_{\rm d}[\hat{\tau}_{\rm d}]$. Otherwise, the receiver takes $\hat{\tau}_{\rm d}$ as the actual peak value position for $s_{\rm d}[n]$. Finally, SOP and coarse CFO estimates can be obtained by computing \eqref{sto} and \eqref{cfo}, respectively. 

To sum up, the proposed double-peak approach for preamble detection and coarse synchronization is formalized in Alg.~\ref{Double_Peak}.

\subsection{Fine Synchronization}
\label{section_fineSync}
After the preamble detection and coarse synchronization succeed, we perform fine synchronization to mitigate fractional timing delay and fractional frequency offset. Specifically, fine timing can be performed within one chip duration around the detected SOP based on an oversampling sequence, i.e., in $\hat{\mu} R + \delta$, $\delta\in [-RK/2, RK/2]$. Then, the oversampling subsequence is decimated by $R$ as $\bar{{\bm r}}_{\delta} = \left[r'[\mu R+\delta], \cdots, r'[\mu R+\delta + (2N - 1) R]\right]$ to reduce the computational complexity.

As the integer part of the CFO, i.e.,  $\Delta f_{\rm I}$, is estimated as per \eqref{cfo} before fine synchronization, the downsampled sequence can be compensated first as $\bar{r}_{\delta}[n] = \bar{r}_{\delta}[n]e^{-j2 \pi \Delta f_{\rm I} n}$. Then, it is filtered with a local reference preamble $s^{\ast}_{\rm pre}[n]$, yielding
\begin{equation} \label{RC}
	\left\vert \Lambda[\delta] \right\vert = \Big|\sum_{n=0}^{2NK-1}\bar{r}_{\delta}[n] \, s^{\ast}_{\rm pre}[n]\Big|,
\end{equation}
where the coarse CFO compensation alleviates much of the correlation peak reduction caused by CFO mismatch. Next, the fine timing is determined by
\begin{equation} \label{tau}
	\hat{\epsilon}_{R} = \hspace{-5pt} \mathop{\arg\max}\limits_{ -{RK}/{2} \leq \delta \leq {RK}/{2}} \hspace{-10pt} \left|\Lambda[\delta]\right|,
\end{equation}
from which we obtain the processed preamble $\tilde{\bm{r}}_{\epsilon}$ after coarse CFO compensation, phase filtering, and timing correction. Finally, the fractional part of the CFO, i.e., $\Delta \hat{f}_{\rm F}$, can be calculated from the autocorrelation angle of $\tilde{\bm{r}}_{\epsilon}$ with a lag of half the preamble length, expressed as
\begin{equation} \label{f_F}
	\Delta \hat{f}_{\rm F} = \frac{1} {2\pi N K} \, \angle\Big\{\sum\limits_{n = 0}^{NK-1} \tilde{r}_{\epsilon}^{\ast}[n] \, \tilde{r}_{\epsilon}[n+N K]\Big\},
\end{equation}
where the phase operator $\angle\left\{\cdot\right\}$ constrains the estimation range of $\Delta f_{\rm F}$ no larger than ${1}/\left(2NK\right)$. Compared to \eqref{f_res}, it is clear that this range is larger than the maximum remaining CFO.

To sum up, the estimated CFO is explicitly determined by the sum of the integer and fractional parts of the CFO, i.e., $\Delta \hat{f} = \Delta \hat{f}_{\rm I} + \Delta \hat{f}_{\rm F}$, which is used to adjust the numerically controlled oscillator for CFO compensation of the remaining header and payload samples. The estimated STO is determined by $\hat{\tau} = \hat{\mu} + \hat{\epsilon}_{R}/R$, and then used to adjust the start index of incoming packets at the receiver.

\begin{algorithm}[!t]
	\caption{\small Preamble Detection and Coarse Synchronization}
	\label{Double_Peak}
	\footnotesize
	\begin{algorithmic}[1]
	\REQUIRE {The received data sequence $r[n]$, the references of down- and up-chirp signals $s_{\rm d}^{\ast}[n]$ and $s_{\rm u}^{\ast}[n]$, the sliding-window lengths $W_{\rm d}$ and $W_{\rm u}$, the scaling factors $\gamma$ and $\eta$;}
	\ENSURE {$\hat{\mu}$ and $\Delta \hat{f}_{\rm I}$;}
	\STATE Initialization: pream\_detection = False, $p = 0$,
	\WHILE{pream\_detection = False}
		\FOR{$k= p_{\rm 1}$ to $p_{\rm 1} + W_{\rm d}-1$}
			\STATE Compute $\left|\rho_{\mathrm{d}}[k]\right|$ as per \eqref{corr},
		\ENDFOR
		\STATE Compute $\hat{\tau}_{\rm d}$ as per \eqref{MLE},
		\IF{$\left|\rho_{\rm d}[\hat{\tau}_{\rm d}]\right| < \gamma \, \mathbb{E}\left(\left|\rho_{\rm d}[k]\right| \right), \, k \notin \left\{\hat{\tau}_{\rm d}-1, \hat{\tau}_{\rm d},\hat{\tau}_{\rm d}+ 1\right\}$} 
			\STATE $p_{1} \gets p_{1}+W_{\rm d}$,
		\ELSE
			\STATE $\hat{\tau}_{\rm c} \gets \hat{\tau}_{\rm d} + NK$,
			\FOR{$k = \hat{\tau}_{\rm c} - W_{\rm u}$ to $\hat{\tau}_{\rm c}+W_{\rm u}$}
				\STATE Compute $\left|\rho_{\rm u}[k]\right|$ in a way similar to \eqref{corr},
			\ENDFOR
			\STATE Compute $\hat{\tau}_{\rm u}$ in a way similar to \eqref{MLE},
			\IF{$\left|\rho_{\rm u}[\hat{\tau}_{\rm u}]\right| < \gamma \, \mathbb{E}\left(\left|\rho_{\rm u}[k]\right| \right), \, k \notin \left\{\hat{\tau}_{\rm u}-1, \hat{\tau}_{\rm u},\hat{\tau}_{\rm u}+ 1\right\}$}
				\STATE $p_{1} \gets p_{1}+W_{\rm d}$
			\ELSE
				\STATE pream\_detection = True
				\IF{$\left|\rho_{\rm u}\left[\hat{\tau}_{\rm u}\right]\right| > \eta\left|\rho_{\rm d}\left[\hat{\tau}_{\rm d}\right]\right|$}
					\FOR{$k = p_{1}+W_{\rm d}$ to $\hat{\tau}_{\rm u}-NK +W_{\rm u}/2$}
						\STATE Compute $\left|\rho_{\mathrm{d}}[k]\right|$ as per \eqref{corr},
					\ENDFOR
					\STATE Compute $\hat{\tau}^{\prime}_{\rm d}$ in a way similar to \eqref{MLE},
					\IF{$\rho_{\rm d}[\hat{\tau}^{\prime}_{\rm d}] > \rho_{\rm d}[\hat{\tau}_{\rm d}]$}
						\STATE $\hat{\tau}_{\rm d} \gets \hat{\tau}^{\prime}_{\rm d}$,
					\ENDIF
				\ENDIF
			\ENDIF
		\ENDIF
	\ENDWHILE
	\STATE{Compute $\hat{\mu} $ and $\Delta \hat{f}_{\rm I}$ as per \eqref{sto} and \eqref{cfo}, respectively.}
\end{algorithmic}
\end{algorithm}

\subsection{Analysis of Cramer-Rao Lower Bound (CRLB)}
We derive the CRLB  to evaluate the CFO estimation quality. In particular, the perfect timing preamble samples can be expressed as
\begin{equation} \label{rx_perfect}
 	r[n] = he^{j\theta_{h}}s_{\rm pre}[n]e^{j 2\pi n \Delta{f}} + w[n], \, n =  0,1,\cdots, 2NK-1
\end{equation}
where the first term on the right-hand-side (RHS) of \eqref{rx_perfect} denotes the desired signal $\bm{s}(\bm{\xi})$ with $\bm{\xi} \triangleq \left[h \ \theta_{h} \ \Delta f\right]^{\rm T}$ being a vector composed of three unknown parameters, and the second term represents AWGN. Therefore, the lower bound on the estimation of the $i^{\rm th}$ parameter in $\bm{\xi}$ is \cite{Estimation}
\begin{equation} \label{crlb1}
	{\rm Var}\left(\hat{\xi}_{i}\right) \geq \left[\bm{I}(\bm{\xi})^{-1}\right]_{i,i},
\end{equation}    
where $\bm{I}(\bm{\xi})$ is the Fisher information matrix of $\bm{s}(\bm{\xi})$, with the $(i, j)^{\rm th}$ element given by
\begin{equation} \label{Fisher}
	\bm{I}(\bm{\xi})_{i, j} = \frac{2}{\sigma^{2}} \Re \left\{\sum_{i = 0}^{2NK-1}
	\frac{\partial s^{*}(n)}{\partial \xi_{i}} 
	\frac{\partial s(n)}{\partial \xi_{j}}\right\}.
\end{equation}
Consequently, the minimum estimation variance of $\Delta{f}$ is
\begin{equation}
	{\rm Var}\left(\Delta \hat{f}\right) 
		\geq \left[\bm{I}(\bm{\xi})^{-1}\right]_{3,3} 
		= \frac{3}{4\pi^2 NK \left(4N^{2}K^2-1\right){\rm SNR}_{h}},  \label{Fisher3}
\end{equation}
where ${\rm SNR}_{h} = |h|^{2}/\sigma^{2}$ denotes the equivalent SNR for frequency-flat fading channels. Clearly, the CRLB in \eqref{Fisher3} depends solely on the parameters $N$, $K$, and ${\rm SNR}_{h}$. In other words, the CRLB of CFO estimation indicates the optimal performance that can be achieved using various preambles of the same length.

\subsection{Payload Demodulation: Non-Coherent Joint Despreading and Demodulation Approach}
In LPWAN systems, receiver sensitivity is a key system-level performance metric that defines the minimum signal power required for reliable decoding. For a given BER or PER requirement, receiver sensitivity is determined by the required SNR, system bandwidth, and processing gain. Specifically, the receiver sensitivity can be calculated as \cite{semtech2015an1200}:
\begin{equation} \label{sensitivity}
S = -174 + 10\log_{10}({\rm OBW}) + {\rm SNR}_{\min} + {\rm NF},
\end{equation}
where the abbreviation ${\rm OBW}$ denotes the occupied bandwidth of payload symbols, which nearly equals the chip rate $B$ for DSSS-MSK signal, and ${\rm SNR}_{\min}$ represents the minimum SNR requirement for the target packet error rate (PER). The last term $\rm NF$ in \eqref{sensitivity} stands for noise figure, which is typically set to $6$ dB for low-cost LPWAN radio front-end \cite{semtech2015an1200}. It is well-known that the minimum SNR required for MSK signals is significantly higher than that for orthogonal FSK signals when using non-coherent receivers. This is due to the narrower bandwidth occupied by MSK signals, which negatively impacts their sensitivity performance \cite{FSK_DSSS, FSK_Enhanced}. Furthermore, when dealing with DSSS-MSK signals, the separation of non-coherent demodulation at the chip level and the soft or hard despreading processes results in an additional loss of SNR.

To address the disadvantages mentioned above, we propose a non-coherent joint despreading and demodulation scheme that operates at the symbol level and is robust against residual synchronization errors, treating one entire DSSS-MSK signal of ${\rm SF}_{\rm p}$ chips as a single symbol. Thus, the locally matched filter for the symbol `1' can be designed as
\begin{align}
s_{\rm DM,1}^{\ast}[n] 
	 &= \exp\left(-j\phi (n;\bm{d})\right)  \nonumber \\
	 &= \exp\Big(-j \pi \hspace{-5pt} \sum_{i = 0}^{{\rm SF}_{\rm p}-1} \hspace{-5pt} d[i] \, q[n-iK]\Big),
\label{CPFSK_MatchFilter}
\end{align} 
where $\bm{d}$ is the DSSS sequence at the transmitter and $d[i] \in \left\{-1, +1\right\}$ is its $i^{\rm th}$ binary code. Likewise, the matched filter for symbol `0' takes the conjugation of ${\rm s}_{\rm DM,0}^{\ast}[n]$.

The non-coherent matched filter output of the $q^{\rm th}$ despreading bit can be written as
\begin{equation} \label{joint}
	z_{i}[q] = \Big|\sum_{n=0}^{K{\rm SF}_{\rm p}-1}r^{(q)}_{\rm pay}[n] \, {s}^{*}_{{\rm DM}, i}[n]\Big|, \ i = 0, 1
\end{equation}
where $r^{(q)}_{\rm pay}[n]$ refers to the $q^{\rm th}$ received DSSS-MSK symbol after synchronization. Then, the simplified soft demodulation output $\lambda[q]$ of the $q^{\rm th}$ despreading bit is
\begin{equation}\label{soft_demodulation}
	\lambda[q] = z_{1}[q] - z_{0}[q].
\end{equation} 
Compared to the hard-decision-based demodulation scheme, the soft-decision information can better exploit the inherent channel coding gain, but requires more quantization bits and increases the computational complexity of the soft decoder. Therefore, the trade-off needs to be addressed in practice.

In principle, the correlation coefficient $\rho_{\rm DM}(\bm{d})$ of all DSSS-MSK symbols spread by sequence $\bm{d}$ determines the error rate performance, computed as
\begin{equation} \label{Eq-Rho}
\rho_{\rm DM}(\bm{d}) = \frac{1}{K {\rm SF}_{\rm p}}\Big|\sum_{n = 0}^{K {\rm SF}_{\rm p}-1} \hspace{-5pt} s_{\rm DM, 1}[n] \, s^{\ast}_{\rm DM, 0}[n]\Big|.
\end{equation}
Clearly, only if $\rho_{\rm DM}(\bm{d}) = 0$,  i.e., the DSSS-MSK symbols `1' and `0' are orthogonal to each other, can one achieve the best receiver performance. Consequently, we have the following proposition to select the optimal spreading sequences for the best non-coherent receiver.\footnote{In contrast, LoRa modulation in continuous-time is not orthogonal in essence, and it becomes orthogonal only for large modulation order, say, larger than $2^7$ \cite{8723130}. This is a key reason why the minimum spreading factor of LoRa modulation is set to $7$. }
\begin{proposition} \label{Proposition-2}
	The value of $\rho_{\rm DM}(\bm{d})$ given by \eqref{Eq-Rho} is zero if the spreading sequence ${\bm{d}}$ satisfies
		\begin{equation}\label{DSSS_ort}
			\sum\limits_{l=0}^{{\rm SF}_{\rm p} -1} (-1)^{l}d[l] = 0.
		\end{equation}
\end{proposition}

\begin{IEEEproof}
	See Appendix~\ref{Appendix-B}.
\end{IEEEproof}

By Proposition~\ref{Proposition-2}, the SNR gain obtained at the receiver can be computed as
\begin{equation} \label{Eq-SNRgain}
	{\rm SNR}^{({\rm SF}_{\rm p})}_{\rm M} = {\rm SNR}_{\rm F} + 10 \log_{10}({\rm SF}_{\rm p}),
\end{equation}
where ${\rm SNR}^{({\rm SF}_{\rm p})}_{\rm M}$ denotes the received SNR gain for DSSS-MSK with ${\rm SF}_{\rm p}$ chip length, and ${\rm SNR}_{\rm F}$ denotes the received SNR gain for orthogonal FSK without spreading. Clearly, the second term on the RHS of \eqref{Eq-SNRgain} implies that our correlation-based scheme achieves the full spreading gain.

In practice, the phase rotation of the received $q^{\rm th}$ symbol is influenced by the accumulated phases of previous symbols, channel fading, and residual CFO. However, phase rotation does not affect the demodulation step, as the proposed demodulation scheme is non-coherent. Thus, our receiver design is inherently robust to potential phase rotation caused by harsh wireless environments.

Finally, by accounting for the residual CFO, the correlation magnitude of payload symbols can be computed as
\begin{align} \label{c_d}
	\rho_{m}(\Delta f_{\epsilon}) 
	&= \frac{1}{K {\rm SF}_{\rm p}}\left\vert \frac{\sin (\pi \Delta f_{\epsilon} K {\rm SF}_{\rm p})}{\sin(\pi \Delta f_{\epsilon})}\right\vert,
\end{align}
where $\Delta f_{\epsilon}$ is the residual CFO normalized by the sampling frequency $f_{s}$. It is not hard to derive that $\rho_{m}(\Delta f_{\epsilon})$ is larger than $0.9$ if $\left\vert \Delta f_{\epsilon}\right\vert < 1/(4K {\rm SF}_{\rm p})$ or $\left\vert \Delta f_{\epsilon}\right\vert < B/(4 {\rm SF}_{\rm p})$. In our simulation experiments to be discussed shortly, we set $B/(4 {\rm SF}_{\rm p})$ as the target bound to measure the estimation quality of the CFO.

\section{Prototype Development and Running Snapshots} \label{PrototypeDevelopment}
The whole transceiver signal processing chain described above has been implemented in the GNU Radio SDR toolkit for real-world performance evaluation. In this section, we elaborate on the implementation details and present some running snapshots of the field-test trials to illustrate the effectiveness of our system design and prototype development.

\begin{figure}[t!]
	\centering
	\includegraphics[width = 0.5\textwidth]{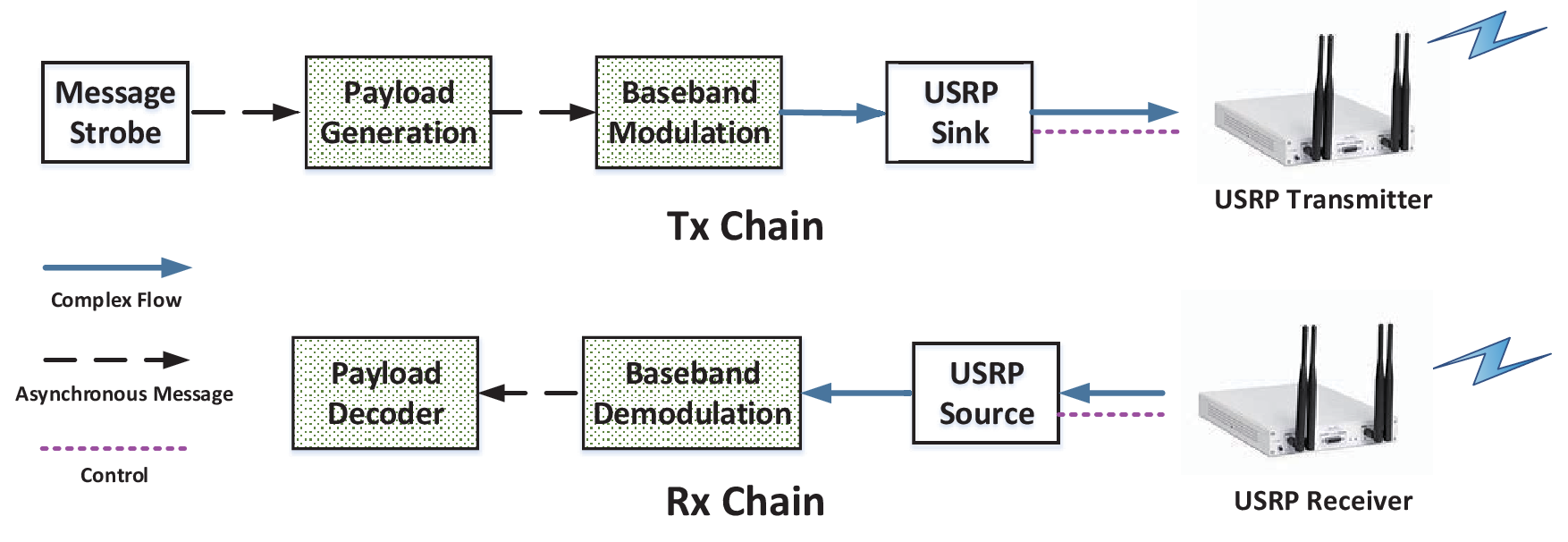} 
	\vspace{-15pt}
	\caption{Transceiver block diagram in GNU Radio.}	
	\label{Transceiver_GNU}
	\vspace{-10pt}
\end{figure}

The prototype development uses general-purpose processors with GNU Radio SDR for baseband signal processing, and Universal Software Radio Peripheral (USRP) Ettus X310 \cite{X310} for RF signal processing, operating at central frequency $f_{c} = 470$ \si{MHz} with a swept-bandwidth $B = 76.8$ \si{kHz}. Figure~\ref{Transceiver_GNU} depicts the transceiver block diagram in GNU Radio SDR, where the {\ttfamily Payload Generation}, {\ttfamily Baseband Modulation}, {\ttfamily Baseband Demodulation}, and {\ttfamily Payload Decoder} blocks are programmed in C++ for performing the specific data processing tasks efficiently. In contrast, the other blocks are derived from GNU Radio SDR. 

\begin{figure}[!t]
	\centerline{\includegraphics[width = 0.5\textwidth]{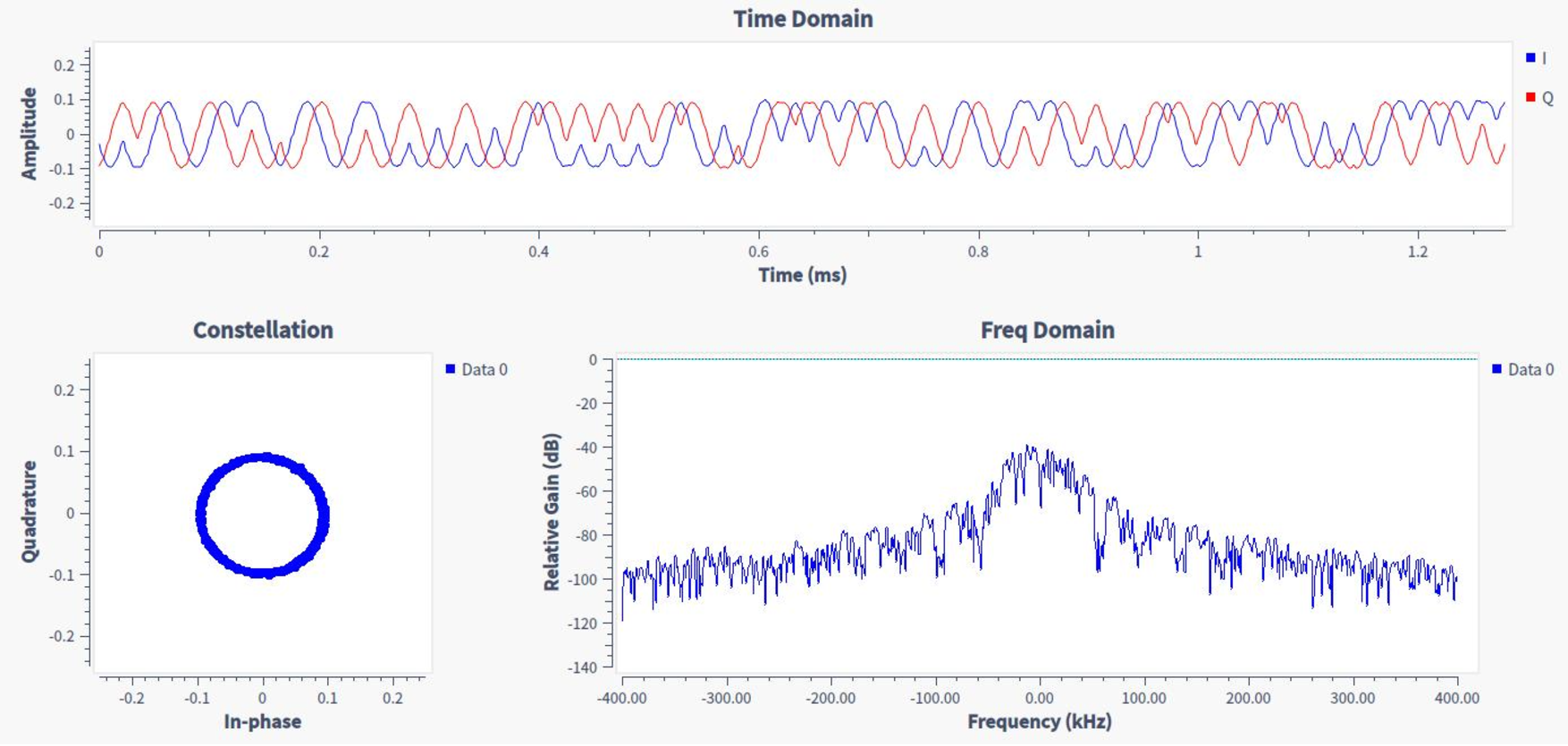}}
	\vspace{-5pt}
	\caption{Three running snapshots of the received payload in the time domain, constellation diagram, and frequency domain.}
	\label{Received1}
	\vspace{-10pt}
\end{figure}

Figure~\ref{Received1} shows three running snapshots of the received payload samples in the time domain, the constellation diagram, and the frequency domain. The payload is a typical MSK signal with a complex-valued Sine waveform in the time domain. The power spectral density (PSD) of the received samples in the frequency domain is clearly of the MSK type. Moreover, the constellation diagram in the lower left corner illustrates the characteristics of constant envelope and phase continuity. 

\begin{figure} [!t]
	\centerline{\includegraphics[width = 0.6\textwidth]{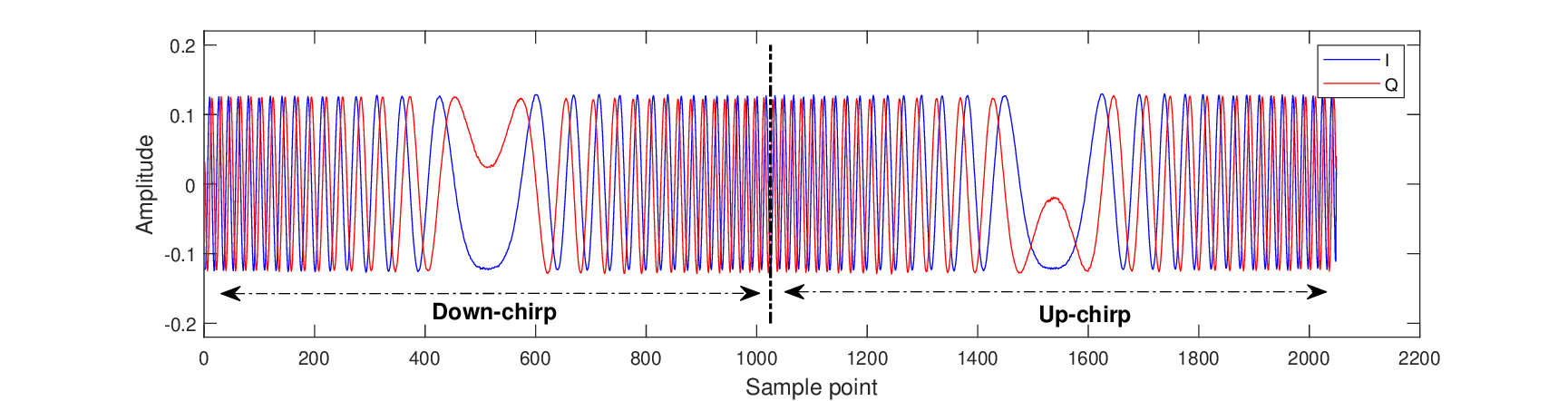}}
	\vspace{-5pt}
	\caption{A running snapshot of the received preamble in the time domain.}
	\label{Received2}
	\vspace{-10pt}
\end{figure}

As the preamble duration in a frame is too short to be captured in a screenshot, we link the {\ttfamily File Sink} block of GNU Radio to the receiver to store the received baseband samples in a separate data file, which are then post-processed and plotted in MATLAB. Figure~\ref{Received2} illustrates that the received preamble is the typical conjugate chirps consisting of linear down- and up-chirps with frequency changes over time. These results demonstrate the effectiveness of our prototype development.

\begin{figure}[t!]
	\centerline{\includegraphics[width = 0.5\textwidth]{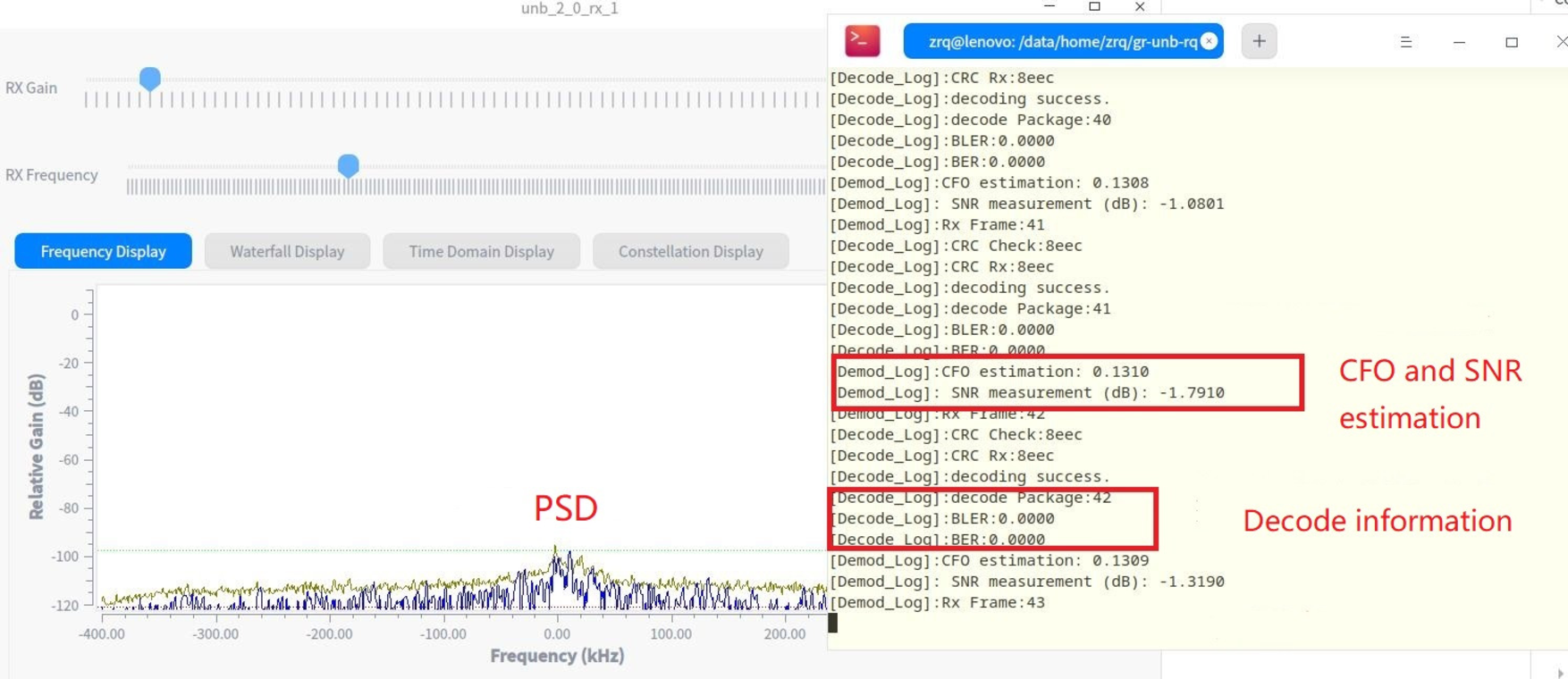}}
	\vspace{-5pt}
	\caption{A running snapshot of the GNU Radio SDR receiver.}
	\label{CFO_est}
	\vspace{-10pt}
\end{figure}

To simulate the potentially high CFO of LPWAN systems in hardware, we intentionally set a $10$ \si{kHz} difference between the Tx and Rx frequencies, yielding a CFO of approximately $13\%$ of the bandwidth $B$. This setting originates from the $f_{c} = 470$ \si{MHz} and the approximate $20$ \si{ppm} precision of a low-cost crystal oscillator. The Tx and Rx gains are also relatively small, simulating the low received SNR of LPWAN systems. Figure~\ref{CFO_est} shows a running snapshot of the SDR receiver, where we highlight the CFO estimation from the {\ttfamily Baseband Demodulation} block. The texts in the red boxes show that the estimates are very close to the manually added values. 

\section{Experimental Results and Discussions} \label{SimulationResults}
This section presents and discusses the Monte Carlo simulation results in MATLAB, comparing them to the field-test results obtained on the GNU Radio SDR-based prototype. 

\subsection{Performance and Complexity of Preamble Detection}
\label{Sync_Compare}
For comparison purposes, two typical preamble detection methods for DSSS-MSK modulation are considered: the double correlation method (referred to as `DSSS A') and the autocorrelation method (referred to as `DSSS B'). The default parameter settings for preamble detection and synchronization are listed in Table~\ref{Table-para1}. Due to the low cost of crystal oscillators, the CFO and STO ranges are intentionally oversized to simulate LPWAN systems with burst communications and high CFOs. For a fair comparison, we define the preamble efficiency as $L_{\rm pre} \triangleq 2N/{\rm SF}_{\rm p}$, which reflects the efficient bit portion in the preamble. The preamble chip length of different detection schemes is set to be identical.

\begin{table}[t!]
\footnotesize
	\renewcommand\arraystretch{1.25}
	\caption{Parameters for Preamble Detection and Synchronization}
	\centering
		\begin{tabular}{!{\vrule width1.2pt} c | c !{\vrule width1.2pt}}
			\Xhline{1.2pt} 
			{\bf Parameter} & {\bf Value} \\
			\Xhline{1.2pt} 
			Chirp length ($N$) & 128 \\
			\hline
			Preamble length ($2N$)& 256 \\
			
			\hline
			Oversampling factor ($K$) &2 \\
			\hline
			Upsampling factor ($R$) & 4 \\
			\hline
			Sliding window width & $NT_{\rm c}$ \\
			\hline
			CFO range & $\mathcal{U}\left(-0.2, 0.2\right) B$ \\   
			\hline                                     
			STO range  & 
			\tabincell{c}{ Case 1: $\mathcal{U}\left(0.25, 0.75\right) NT_{\rm c}$ \\ Case 2: $\mathcal{U}\left(0.25, 1.5\right) NT_{\rm c}$ } \\                    
			\hline
			SNR per symbol & $\left[-10, 10\right]$ \si{dB} \\
			\Xhline{1.2pt} 
		\end{tabular}
		\label{Table-para1}
		\vspace{-10pt}
\end{table} 

\subsubsection{\underline{Preamble Detection Performance}}
In our Monte-Carlo simulation experiments, a frame is detected successfully if its SOP is detected within one chip duration. We first simulate the miss detection rate $P_{\rm M}$ (computed as $1-P_{\rm D}$, where $P_{\rm D}$ represents the detection rate) and false alarm rate $P_{\rm FA}$ of our proposed double-peak detection approach under AWGN channels, with a particular detection threshold $\gamma$.  In each simulation trial under different SNRs, the STO range is set as specified in Case 1 of Table~\ref{Table-para1}. The pure noise samples are generated for $P_{\rm FA}$ calculation and then added to the signal samples affected by STO and CFO for $P_{\rm M}$ calculation. 

\begin{figure}[t!]
	\centering
	\subfloat[\scriptsize $P_{\rm M}$ and $P_{\rm FA}$ under AWGN channels (STO setting follows Case 1 in Table~\ref{Table-para1}.)]{\includegraphics[width=0.45\textwidth]{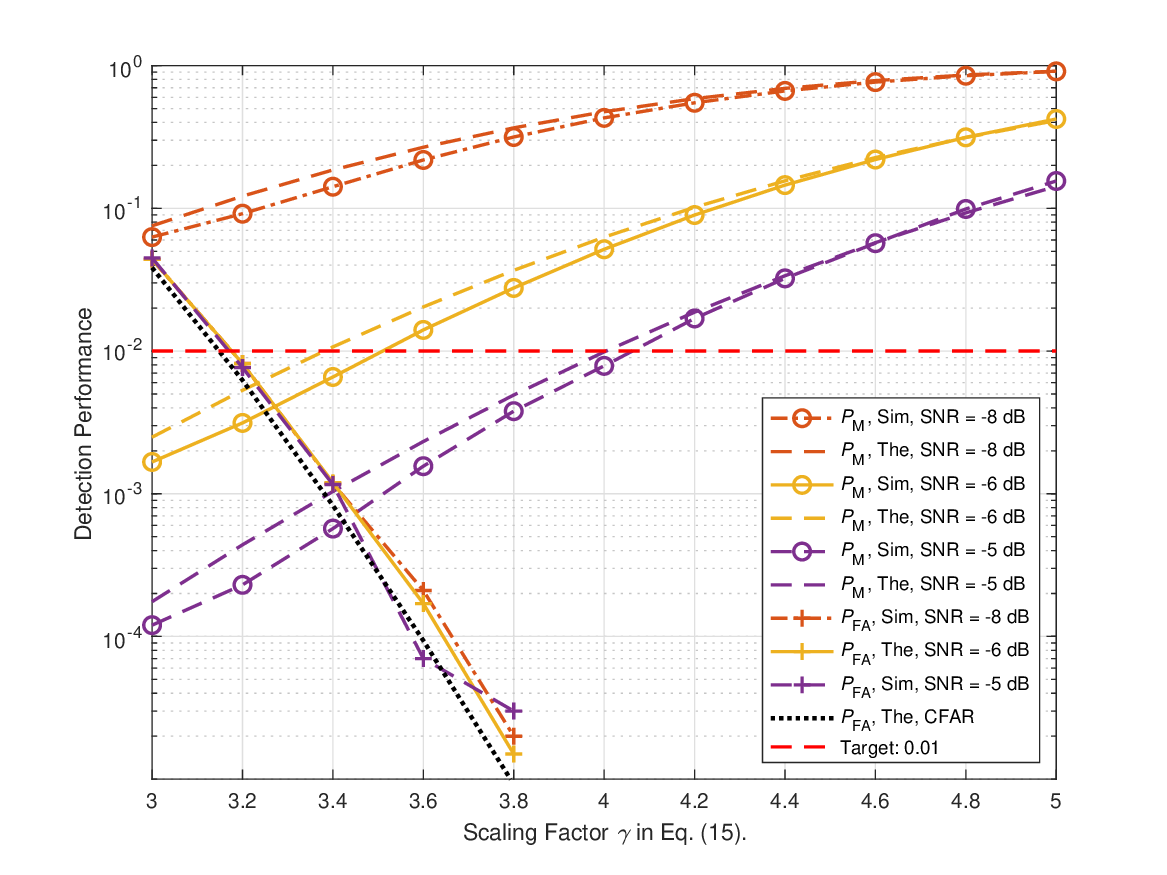} \label{Compare_Detection0}}	
	\hfil
	\subfloat[\scriptsize $P_{\rm D}$ under AWGN/Rayleigh channels (STOs setting follow Case 2 in Table~\ref{Table-para1}).]{\includegraphics[width=0.45\textwidth]{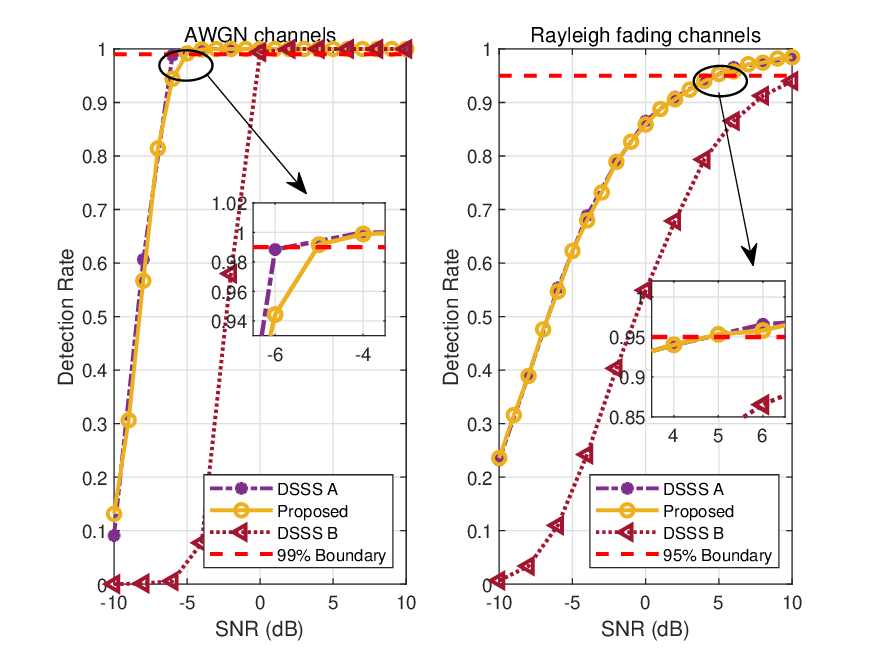} \label{Compare_Detection}}	
	\hfil
	\caption{Preamble detection performance.}	
	\label{Fig-9}
	\vspace{-10pt}
\end{figure}

Figure~\ref{Compare_Detection0} depicts the simulation results with ${\rm SNRs}$ being $-8$, $-6$ and $-5$ \si{dB}, compared to the theoretical results of $P_{\rm FA}$ and $P_{\rm M}$ derived in Appendix~\ref{Appendix-A}. It is observed that the simulation results align well with the theoretical ones, particularly for a large $\gamma$ setting with respect to $P_{\rm M}$ and a small $\gamma$ setting for $P_{\rm FA}$. Based on the $P_{\rm M}$ and $P_{\rm FA}$ curves, we set $\gamma = 4$ for extremely low $P_{\rm FA}$ (i.e., $P_{\rm FA} < 10^{-5}$) and relatively low $P_{\rm M}$ in the following MATLAB simulations and prototype tests, which still reaches the target $P_{\rm M} = 0.01$ at ${\rm SNR} = -5$ \si{dB}. 

Next, we simulate the preamble detection rate $P_{\rm D}$ of each scheme under AWGN channels (i.e., $|h| \equiv 1$) and frequency-flat Rayleigh fading channels (i.e., $h \sim \mathcal{CN}(0,1)$ for every received frame). The STO setting follows Case 2 for the more common situation where the actual frame arrival time is unknown at the receiver. In this case, the down-chirp may not appear in the first sliding window, and the receiver will continue to detect the received samples. The target $P_{\rm D}$ is set to $99\%$ and $95\%$ under AWGN and Rayleigh channels, respectively, to ensure the overall transceiver quality. 
 
Figure~\ref{Compare_Detection} shows that our detection method performs almost the same as the DSSS A scheme, no matter under AWGN or Rayleigh channels, which reaches the target $P_{\rm D}$ $99\%$ at ${\rm SNR} = -5$ \si{dB} under AWGN channels, or $95\%$ at ${\rm SNR} = 5$ \si{dB} under Rayleigh channels. In particular, $P_{\rm D}$ reaches $100\%$ under AWGN channels when SNR $> -4$ \si{dB} because the criterion of \eqref{Detection2} improves the robustness of the preamble detection at high SNR. The reason our approach performs similarly to the DSSS A scheme is that the method for computing the cross-correlation of down- and up-chirps is similar to the double correlation of DSSS codes. However, the chirp signals have higher CFO-tolerance capability, as previously demonstrated in Section~\ref{Section-Chirp_analysis}. The DSSS~B method performs the worst, yielding more than $5$ \si{dB} performance degradation under both AWGN and Rayleigh channels.

\begin{table}[!t]
\footnotesize
	\renewcommand\arraystretch{1.25}
	\caption{Complexity of Different Preamble Detections}
		\vspace{-5pt}
	\centering
		\begin{tabular}{!{\vrule width1.2pt} c | c !{\vrule width1.2pt}}
		\Xhline{1.2pt} 
		\textbf{Method} & \textbf{Computational Complexity} \\
		\Xhline{1.2pt} 
		DSSS A &  $\mathcal{O}(8N^3K^3)$ \\
		\hline
		DSSS B& $\mathcal{O} (4N^{2}K^{2})$\\	
		\hline 
		Proposed & \tabincell{c}{ $\mathcal{O}(4N^2K^2)$, w/o FFT\\ $\mathcal{O}(2NK\log_{2}(2NK))$, w/ FFT}  \\
		\Xhline{1.2pt} 
	\end{tabular}
	\label{Table-complexity}
	\vspace{-10pt}
\end{table} 

\subsubsection{\underline{Complexity Analysis}}
As mentioned in Section~\ref{Preamble_Detection}, the coarse STO and CFO estimation can be jointly performed by the double-peak detection approach, whose computational complexity is generally of order $\mathcal{O} \left((2N)^2K^2 \right)$ for the cross-correlation calculation. If the FFT-based fast correlation approach is applied to replace the cross-correlation approach, the computational complexity can be further reduced to $\mathcal{O} \left(2NK\log_{2}(2NK) \right)$. In comparison, the DSSS B method achieves the same order for the preamble detection/coarse STO estimation as $\mathcal{O} \left((2N)^2K^2 \right)$ because it performs autocorrelation only once on the filtered samples. DSSS~A method combines the autocorrelation magnitudes with lags from $1$ to $D$ of the filtered samples for accurate estimation at low SNR, where $D$ is related to the preamble length. As a result, its computational complexity is of order $\mathcal{O} \left((2N)^3K^3\right)$, which is the highest of these three detection methods.

In summary, Table~\ref{Table-complexity} compares the complexity of these schemes, revealing that the proposed method achieves the lowest computational complexity. Last but not least, it is noteworthy that both DSSS-based preamble detection methods cannot calculate coarse CFO and require additional operations during the synchronization step, such as FFT-based estimation \cite{MSK_Sync}. This adds to the computational overhead. Therefore, using two conjugate base chirps instead of the traditional DSSS-based preamble in our streamlined preamble structure, we can significantly reduce the computational overhead in preamble detection while maintaining satisfactory performance.

\subsection{Synchronization Performance} 
The CFO and STO estimation performance is evaluated under AWGN channels using the mean squared error (MSE). By definition, the MSE is computed as $\mathbb{E}\{\left|\hat{x} - x\right|^{2}\}$, where $x$ denotes an actual normalized synchronization parameter and $\hat{x}$ the estimate. The simulation algorithm follows the same steps as Section \ref{Preamble_Detection}, except that it omits the threshold judgment step related to \eqref{PAR} and \eqref{Detection2} to minimize the impact of packet detection failure on estimation quality. As a result, we set STO as per Case 1 in Table~\ref{Table-para1} to ensure that the preamble is detected in the first sliding window.

\begin{figure}
	\centering
	\includegraphics[width = 0.45\textwidth]{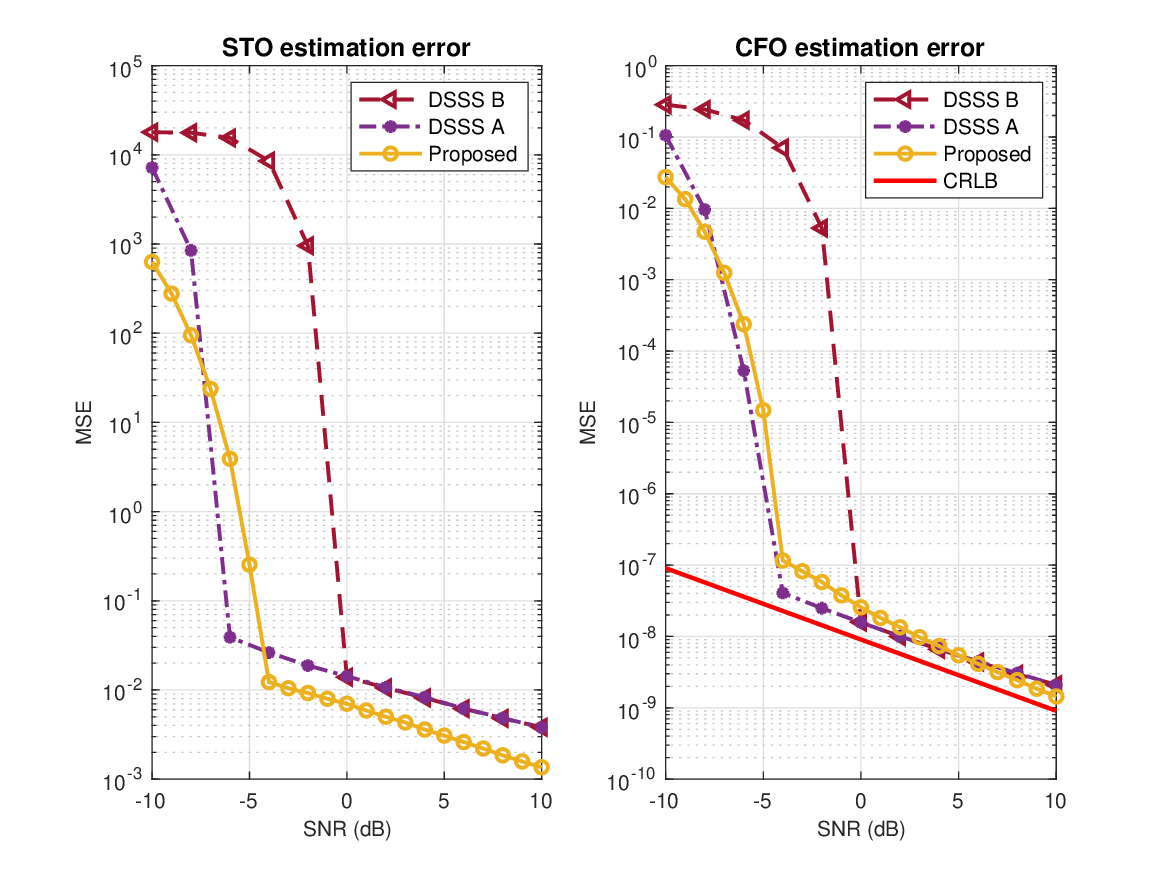}
	\vspace{-5pt}
	\caption{The MSE of STO/CFO estimation under AWGN channels, where STOs are set as PER Case 1 in Table~\ref{Table-para1}.}
	\vspace{-10pt}
	\label{Compare_Sync}
\end{figure}

The left panel of Figure~\ref{Compare_Sync} illustrates the MSE of STO estimation of different methods. It is seen that the DSSS~B performs the worst and gets synchronized only if the SNR is greater than $0$ \si{dB}. This observation agrees with that in the left panel of Figure~\ref{Compare_Detection}, where the preamble detection rate becomes $100\%$ only if $\rm{SNR} > 0$ \si{dB}. The DSSS A method performs similarly to the DSSS B when $\rm{SNR} > 0$ \si{dB}, as they differ only in the packet detection steps. In other words, they have the same synchronization performance if they both have perfect packet detection. On the other hand, it is seen that the proposed method achieves the best STO estimation when $\rm{SNR} > -4$ \si{dB}, thanks to the powerful CSS used in our preamble. In contrast, in the case of $\rm{SNR} < -4$ \si{dB}, the proposed scheme underperforms DSSS A, due to the peak leakage effect discussed after \eqref{correlation}.

The right panel of Figure~\ref{Compare_Sync} compares the MSE of CFO estimation of different methods. We observe that different synchronization methods can approach the CRLB of CFO estimation at high SNR. At low SNR, the DSSS B method has poor coarse timing performance. It is hard to align with the received preamble, resulting in a significant CFO estimation error until ${\rm SNR} > 0$ \si{dB}, when it coincides with the performance curve of the DSSS A. Chirp signals have a peak leakage problem when doing correlation detection in the digital domain, so the deviation of the coarse CFO estimation may be significant, which is hard to compensate for in the fine CFO estimation step. Therefore, when ${\rm SNR} < 3$ dB, the CFO estimation performance of the proposed method is slightly worse than that of the DSSS A. It is not until ${\rm SNR} > 3$ dB that the performance of the proposed method becomes better.

\subsection{Payload Demodulation Performance} 
\label{PER_Performance}
Packet error rate (PER), defined as the ratio of the number of unsuccessfully received packets to the total number of transmitted ones, is one of the most critical performance indicators of an end-to-end transceiver. In this subsection, we first simulate the PER performance in MATLAB to obtain the lower bound of the proposed payload demodulation scheme and evaluate whether our synchronization scheme limits the overall PER performance. Then, we compare the field-test results with the Monte Carlo simulation results to illustrate the effectiveness of our transceiver design and prototype development. In the related experiments, the payload length is set to $50$ bytes and the spreading factor ${\rm SF}_{\rm p} \in \left\{4,8,16\right\}$. Also, we adopt the same $(2, 1, 7)$ convolutional code as in IEEE LPWAN \cite{IEEE802154}.

\subsubsection{\underline{Simulation Results in MATLAB}}
\label{PER_1}
\begin{figure}[t!]
	\centering
	\subfloat[\scriptsize PER under AWGN and Rayleigh channels in the case of ideal synchronization.]
	{\includegraphics[width=0.45\textwidth]{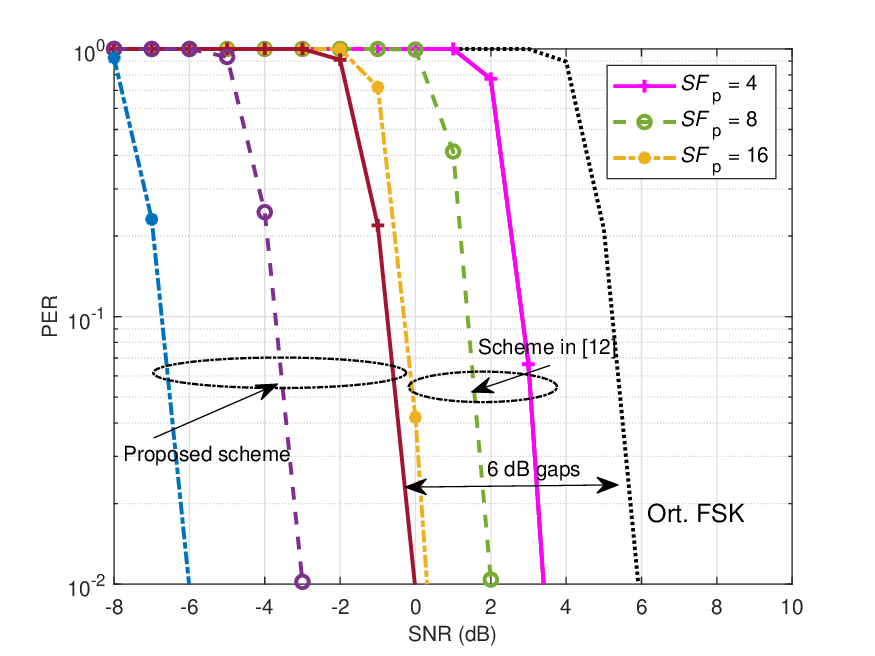} 
	\label{Compare_BLER0}}	
	\hfil
	\subfloat[\scriptsize Overall PER: actual preamble detection and synchronization vs. ideal ones.]
	{\includegraphics[width=0.45\textwidth]{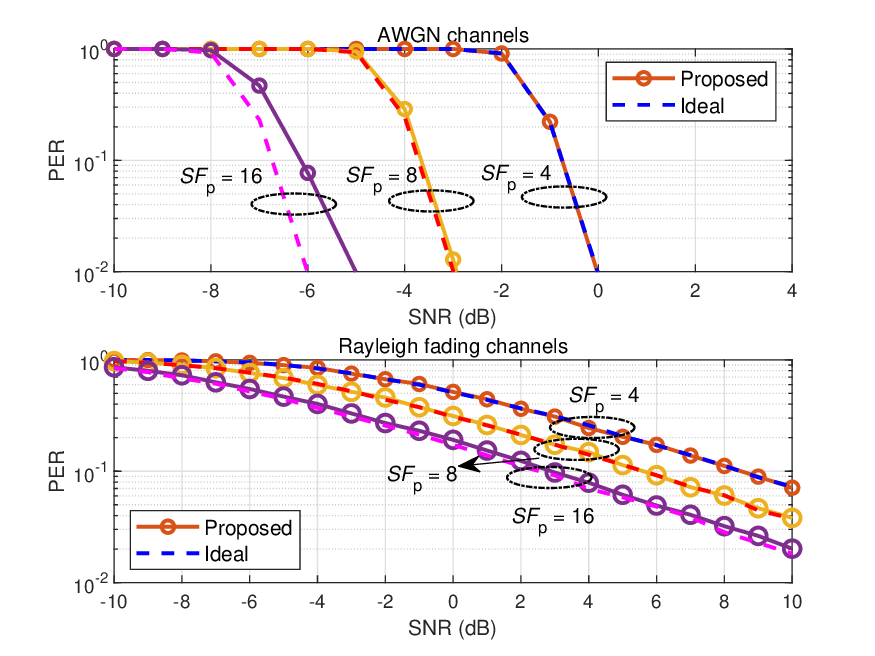} 
	\label{Compare_BLER}}	
	\hfil
	\caption{Payload demodulation performance.}	
	\label{Fig-11}
	\vspace{-10pt}
\end{figure}

Figure~\ref{Compare_BLER0} compares the PER performance of our proposed joint despreading and demodulation scheme with that in \cite{xu2017design} under AWGN channels, assuming ideal synchronization. It can be observed that given the target $ {\rm PER} = 0.01$, the SNR requirement of our scheme is about $\left\{0, -3, -6\right\}$ \si{dB} with 3 \si{dB} spreading gains if ${\rm SF}_{\rm p}$ is doubled, whereas for the scheme in \cite{xu2017design}, it needs  $\left\{3.5, 2, 0.2\right\}$ with no more than $2$ \si{dB} spreading gains if ${\rm SF}_{\rm p}$ is doubled. The SNR requirement for orthogonal FSK detection is $6$ \si{dB}, resulting in a $6$ \si{dB} gap with our proposed scheme for DSSS-MSK signal detection when ${\rm SF}_{\rm p} = 4$. Therefore, it can be inferred that the DSSS-MSK signals turn orthogonal if the spreading sequence $\bm{d}$ satisfies Proposition~\ref{Proposition-2}. In real-world applications, these SNR gains can be leveraged to reduce the spreading sequence length or decrease the transmit power, enabling more economical and power-efficient LPWAN transceivers. 
 
Next, we evaluate the influence of preamble detection and synchronization on the overall PER performance, where the synchronization parameter setting is the same as Table \ref{Table-para1}, and STO is set as per Case 2 in Table~\ref{Table-para1}. The preamble efficiency $L_{\rm pre} = 64, 32, 16$ for ${\rm SF}_{\rm p} = 4, 8, 16$ given $N = 128$, respectively. Figure~\ref{Compare_BLER} shows the overall PER performance under AWGN and Rayleigh channels, where the PER curves are compared to their lower bounds under different channels and ${\rm SF}_{\rm p}$ values. Note that the lower bounds are obtained in the case of ideal synchronization. It is observed that the proposed scheme approaches the lower bounds in ${\rm SF}_{\rm p} = 4, 8$ under both channels. In contrast, the proposed scheme yields about $1$ \si{dB} degradation if ${\rm SF}_{\rm p} = 16$ under AWGN channels, which is mainly because the preamble length is not long enough to reach `good synchronization' in the relatively low SNR region. Under the Rayleigh channels with ${\rm SF}_{\rm p} = 16$, the proposed scheme still aligns with the lower bound, primarily because the channel fading affects both the preamble synchronization and payload demodulation simultaneously.

The above results indicate that a preamble length of $L_{\rm pre} = 32$ is sufficient to achieve the desired PER performance in practice. The setting $L_{\rm pre} = 16$ is not long enough and degrades PER, whereas $L_{\rm pre} = 64$ wastes the preamble since the SNR region for good synchronization is much lower than for payload demodulation. These observations are crucial for selecting parameters during the development of the accompanying prototype.

\subsubsection{\underline{Field-test Trial Results}}
Now, we present and discuss the PER of field-test trials. The default RF parameters are summarized in Table~\ref{Table_para3}, and PHY parameters are identical to those used before. In our testbed deployment, an open aisle exists between the two USRPs, allowing for a line-of-sight path between the transmitter and the corresponding receiver. This results in a nearly time-invariant channel response during our experimental test, enabling PER measurements to be performed under quasi-AWGN channels. We fix the $G_{\rm Tx}$ and sweep the $G_{\rm Rx}$ with a $1$ \si{dB} gap to generate different received SNR values for prototype tests. To simulate the influence of potentially high CFOs, in the CFO-added case, we intentionally set a 10 \si{kHz} difference between the $f_{\rm Tx}$ and $f_{\rm Rx}$, yielding a CFO of about $13\%$ of $B$. In contrast, they are identical in the CFO-free case. To further demonstrate the effectiveness of our prototype development, we compare field-test results with MATLAB simulation ones, given the same parameter settings and lower bounds of PER generated in Subsection~\ref{PER_1}.

\begin{table}[!t]
	\footnotesize
	\caption{RF Parameters in Field-test Trials}
	\vspace{-5pt}	
	\centering
	\renewcommand\arraystretch{1.25}
		\begin{tabular}{!{\vrule width1.2pt} c | c !{\vrule width1.2pt}}
		\Xhline{1.2pt} 
		{\bf Parameter} & {\bf Value} \\
		\Xhline{1.2pt} 
		Tx frequency ($f_{\rm Tx}$) & 470 \si{MHz} \\
		\hline
		Rx frequency ($f_{\rm Rx}$) & 470 \si{MHz} (+10 \si{kHz}) \\
		\hline
		Bandwidth/Chip rate ($B$) & 76.8 \si{kHz} \\		
		\hline
		Tx gain ($G_{\rm Tx}$) & 0 \si{dB} \\
		\hline
		Rx gain ($G_{\rm Rx}$) & $[0, 9]$ \si{dB} \\
		\Xhline{1.2pt} 
	\end{tabular}
	\label{Table_para3}
	\vspace{-10pt}
\end{table} 

\begin{figure}
	\centering
	\includegraphics[width=0.45\textwidth]{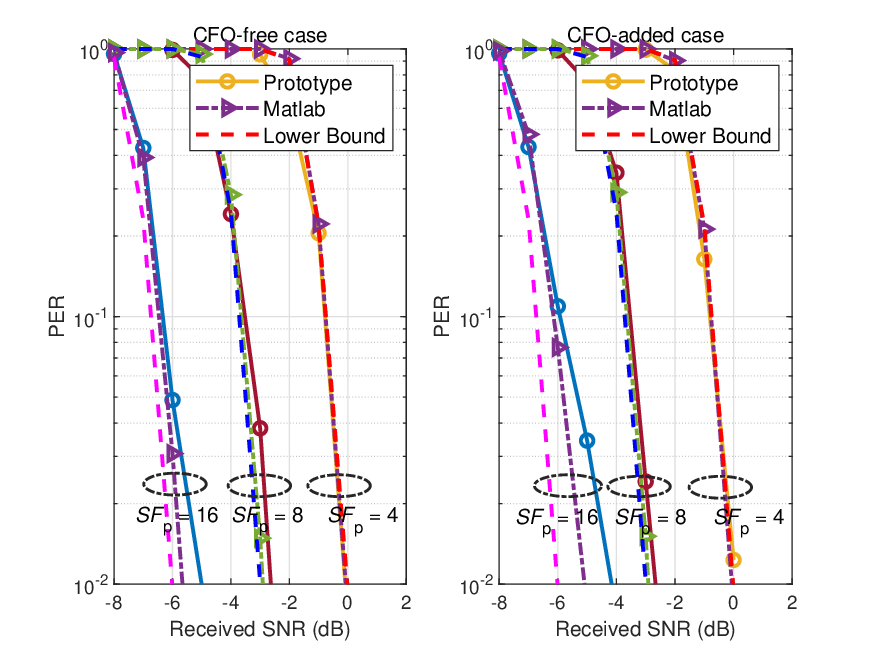}
	\vspace{-10pt}	
	\caption{PER comparison of field-tests vs. MATLAB simulation results.}
	\vspace{-10pt}
	\label{BLER_Compare}
\end{figure}

Figure~\ref{BLER_Compare} plots the PER curves of field-test trials in the CFO-free and CFO-added cases, respectively. Regardless of the prominent added CFOs, the field-test results coincide with the MATLAB simulation ones for ${\rm SF}_{\rm p} = 4, 8$, demonstrating the effectiveness of our prototype development. In the case of ${\rm SF}_{\rm p} = 16$, there exist gaps between the PER lower bounds and the MATLAB simulations and the field-test results. This is because the preamble is not sufficiently long, as mentioned above. On the other hand, the CFO-free results outperform the CFO-added ones for both the MATLAB simulations and prototype trials without any surprise because the correlation peak diminishes due to the extensive added CFO. Finally, it is not surprising that the results of prototype trials perform slightly worse than those of MATLAB simulation experiments due to hardware implementation losses.

\subsection{Comparison of Four Typical LPWAN PHYs}
For engineering purposes, we compare our proposed LPWAN PHY with the NB-IoT PHY, LoRa PHY, and an enhanced FSK PHY designed in \cite{FSK_DSSS}, based on their publicly available parameters. We take ${\rm SF}_{\rm p} = 8$ as the benchmark spreading factor, where ${\rm SNR}_{\min} = -3$ \si{dB} for 50-byte packet length, and nearly $-122$ \si{dBm} receiver sensitivity can be achieved according to \eqref{sensitivity}. Given coding rate ${\rm CR}_{\rm p} = 1/2$ and spreading factor ${\rm SF}_{\rm p} = 8$, its effective bit rate is ${\rm CR}_{\rm p} \times B /{\rm SF}_{\rm p} = 4.8$ \si{kbps}, which serves as the data rate benchmark. For a fair comparison, the receiver sensitivity of different technologies is compared under almost the same effective data rate, as detailed below.

\subsubsection{\underline{NB-IoT PHY}} By Release 15 of 3GPP \cite{36101}, its reference sensitivity is $S_{\rm NB} = -108.2$ \si{dBm} in the case of coding rate ${\rm CR}_{\rm NB} = 1/3$, no repetition, and quadrature phase shift keying (QPSK) constellation carrying $\gamma_{Q} = 2$ bits per symbol. The maximum throughput of NB-IoT is $R_{\rm NB} = 320$ \si{kbps} in the absence of coding and repetitions.\footnote{Each NB-IoT subframe with $1$ \si{ms} duration is divided into 14 OFDM symbols and has $N_{\rm sub} = 12$ subcarriers in the frequency domain. Thus, $N_{\rm g} = 14 \times N_{\rm sub} - 8 = 160$ time-frequency grids can be used in one subframe as $8$ grids are filled with reference signals. Therefore, if QPSK is adopted, the number of transmitted raw bits per second is $N_{\rm g} \gamma_{Q} \times 1000 = 320000$.} Thus, to reduce the effective bit rate, we set the repetition times $N_{\rm rep} = 32$, which changes the effective bit rate to $ {\rm CR}_{\rm NB}/N_{\rm rep} \times R_{\rm NB} \approx 3.33$ \si{kbps}. For ideal conditions, we assume the full repetition gains can be achieved and the modified sensitivity is $S'_{\rm NB} = S_{\rm NB} - 10\log_{10}(N_{\rm rep}) \approx -123.2$ \si{dBm}.
	
\subsubsection{\underline{LoRa PHY}} By the SX1272 data sheet \cite{sx1272} provided by Semtech, the LoRa CSS sensitivity is $-124$ \si{dBm} given the spreading factor ${\rm SF}_{\rm LR} = 7$, the bandwidth $B_{\rm LR} = 125$ \si{kHz}, and the modified Hamming code with coding rate ${\rm CR}_{\rm LR} = 4/6$, yielding its effective bit rate being ${\rm CR}_{\rm LR} \times {\rm SF}_{\rm LR} / 2^{{\rm SF}_{\rm LR}} \times B_{\rm LR} = 4.56$ \si{kbps}.
	
\subsubsection{\underline{FSK PHY \cite{FSK_DSSS}}} The $12.5 k$ rate mode (i.e., $B_{\rm DS} = 12.5$ \si{kHz}) in Table~I of \cite{FSK_DSSS} needs ${\rm SNR} = 13$ \si{dB} to decode received packets, without channel coding and DSSS. To keep the effective bit rate at the same level as our comparison benchmark, we set the DSSS length ${\rm SF}_{\rm DS} = 4$ so that the effective bit rate changes to $B_{\rm DS}/{\rm SF}_{\rm DS} = 3.125$ \si{kbps}. Due to the significant SNR loss caused by the hard-decision-based despreading scheme, the ${\rm SNR}_{\min}$ only decreases to $11$ \si{dB} given ${\rm SF}_{\rm DS} = 4$. The FSK occupied bandwidth of this rate mode is $50$ \si{kHz}; thus, the receiver sensitivity is $-110$ \si{dBm} as per \eqref{sensitivity}.

In summary, Table~\ref{Table_FourSchemes} compares four schemes regarding receiver sensitivity and effective bit rate. Thanks to the more advanced transceiver design, our proposed transceiver outperforms the low-cost FSK PHY of \cite{FSK_DSSS} in terms of receiver sensitivity and effective bit rate. Compared to the commercial NB-IoT and LoRa solutions, the receiver sensitivity of the proposed scheme is similar. Still, the proposed DSSS-CPFSK PHY enables reduced complexity and cost of the transceiver. In particular, replacing the traditional DSSS-based preamble with a conjugate chirp-based preamble significantly reduces the computational complexity and power consumption at the preamble detection step, enabling more power-saving and cost-effective LPWAN implementations.

\begin{table}[t!]
	\small
	\renewcommand\arraystretch{1.25}
	\caption{Comparison of Four Typical LPWAN Schemes}
	\vspace{-5pt}
	\centering
	\begin{tabular}{!{\vrule width1.2pt} c !{\vrule width1.2pt} c | c | c | c !{\vrule width1.2pt}}
		\Xhline{1.2pt} 
		\hline
		& {\footnotesize Proposed} &{\footnotesize NB-IoT PHY} & {\footnotesize LoRa PHY} & {\footnotesize FSK PHY} \\
		\Xhline{1.2pt} 
		{\footnotesize Sensitivity} & \footnotesize -122 \si{dBm} & \footnotesize -123.2 \si{dBm} & \footnotesize -124 \si{dBm} & \footnotesize -110 \si{dBm} \\
		\hline
		{\footnotesize Data~rate}& \footnotesize 4.8 \si{kbps} & \footnotesize 3.33 \si{kbps} & \footnotesize 4.56 \si{kbps} & \footnotesize 3.125 \si{kbps} \\
		\Xhline{1.2pt} 
	\end{tabular}
	\label{Table_FourSchemes}
	\vspace{-10pt}
\end{table} 

\section{Concluding Remarks} 
\label{ConcludingRemarks}
In this paper, we have designed and prototyped an end-to-end LPWAN transceiver that features a concise preamble, comprising a pair of conjugate down- and up-chirps, and a DSSS-CPFSK payload modulation scheme. At the receiver, we introduced a double-peak preamble detector, enabling low-complexity coarse and fine synchronization that is robust to carrier frequency offset (CFO) and phase rotation. For payload recovery, we developed a non-coherent joint despreading and demodulation scheme, which achieves lower complexity than conventional coherent receivers. Compared to commercial LoRa and NB-IoT solutions, the proposed transceiver offers similar receiver sensitivity and higher effective bit rate, while maintaining lower receiver and computational complexity. These features make the design particularly suitable for IoT applications with relatively static or semi-static channels, such as those found in the metering industry (e.g., smart electricity, water, gas, and heat meters). This work focuses on physical-layer transceiver design and link-level performance evaluation. System-level aspects, including MAC-layer interactions, access coordination, and network scalability, are important extensions and are left for future study.

\begin{appendices}
\section{Proof of Proposition~\ref{Proposition-1}} \label{Appendix-A}

During one sliding window, the correlation $|\rho_{\rm d}[k]|$ given by \eqref{corr}, $k = p_{1}, \cdots,p_{1}+W_{\rm d}-1$, are distributed as Rayleigh distribution under null hypothesis $\mathcal{H}_{0}$, or approximate Rayleigh-distributed for $ k \notin \{\hat{\tau}_{\rm d}-1, \hat{\tau}_{\rm d}, \hat{\tau}_{\rm d}+1\}$ under alternative hypothesis $\mathcal{H}_{1}$. Thus, we have $\left|\rho_{\rm d}[k]\right| \sim {\rm Rayleigh(\sigma_{\rm MF})}$, where $\sigma_{\rm MF} = \sigma /\sqrt{2 N}$ is the equivalent noise standard deviation after matched filtering \cite{Chirp_Sync1}. Then, by \cite[Eq. (2.22)]{Detection}, the expectation of $\left|\rho_{\rm d}[k]\right|$ can be approximated as $\mathbb{E}\left[\left|\rho_{\rm d}[k]\right| \right] \approx \sqrt{\pi/2} \, \sigma_{\rm MF}$. Therefore, the probability of $|c[k]|$ smaller than the threshold can be computed as
\begin{equation} \label{properity}
	p  = \int_{0}^{\gamma \mathbb{E}\left[\left|\rho_{\rm d}[k]\right| \right]} \hspace{-5pt} \frac{x}{\sigma_{\rm MF}} \exp\left(-\frac{x^2}{\sigma_{\rm MF}^2}\right) {\rm d}x 
	 \approx 1-\exp\left(-\frac{\pi}{4} \gamma^2 \right), \nonumber
\end{equation}
which is a constant for a given $\gamma$. Since $\left|c[k]\right|$, $k = p_{1}, \cdots,p_{1}+W_{\rm d}-1$, can be treated as i.i.d., the probability for each $|c[k]|$ in a sliding window smaller than the threshold is approximately $\left(1-\exp\left(-\pi \gamma^2/4\right)\right)^{\rm W }$ with ${\rm W} = {\rm W}_{\rm d} = {\rm W}_{\rm u}$ being the sliding window width. Therefore, the false alarm rate for down-chirp detection can be computed as
\begin{equation}\label{false_alarm}
	P_{\rm FA,d} \approx 1 - \left(1-\exp\left(- \frac{\pi}{4} \gamma^2 \right)\right)^{\rm W }.
\end{equation}

On the other hand, the peak magnitude in sliding window under $\mathcal{H}_{1}$ is of Rician distribution, i.e., $\left | \rho_{\rm d}\left[\hat{\tau}_{\rm d}\right] \right| \sim 
{\rm Rician}\left(m\left(\Delta f\right), \sigma_{\rm MF}\right)$, where $m\left(\Delta f\right)$ is the maximum correlation magnitude affected by CFO $\Delta f$ and channel response $h$, expressed as
\begin{equation}\label{m_f}
	m(\Delta f) = \frac{1}{NK}\left| h\frac{ \sin \left(\frac{\pi \Delta{\epsilon}}{NK^{2}} \left(N K \left(1 - \left|K \Delta f\right| \right) - \Delta{\epsilon} \right)\right)} {\sin \left( \frac{\pi \Delta{\epsilon}}{NK^{2}} \right)} \right|, \nonumber
\end{equation}
in which $\Delta{\epsilon} = \left|\lceil \Delta f K\rfloor - \Delta f K\right|$ is the interval between the actual peak-value position limited by the sampling rate and the ideal one, which causes the peak leakage. Therefore, the down-chirp detection probability for a certain $\Delta f$ is
\begin{align}
	P_{\rm D,d}(\Delta f) &= \int_{\gamma \mathbb{E}[\left|c[k]\right|]}^{+\infty} \frac{x}{\sigma_{\rm MF}^2} \left(-\frac{x^2+m\left(\Delta f\right)^2}{2\sigma_{\rm MF}^{2}}\right) \nonumber \\
	 &\hspace{1em} \times{} I_{\rm 0}\left(\frac{m(\Delta f)x}{\sigma_{\rm MF}^2}\right) {\rm d}x  
	 \approx Q_{{\chi}^{\prime 2}_{2}(\lambda(\Delta f))}\left(\frac{\pi \gamma^2}{2}\right), \nonumber 
\end{align} 
where $ Q_{{\chi}^{\prime 2}_{2}(\lambda(\Delta f))}(\cdot)$ denotes the right-tail probability of non-central Chi-squared random variables with $2$ degrees of freedom \cite[p. 28]{Detection} and $\lambda(\Delta f) = m\left(\Delta f\right)^2/\sigma_{\rm MF}^2$ being the non-central coefficient. To simplify the analysis, we only consider the AWGN case with $|h| \equiv 1$. 

The up-chirp detection will be triggered after the down-chirp detection exceeds the threshold. Since the same correlation method is adopted with the same chirp and detection window widths, it is straightforward that they have the same detection and false alarm rates, i.e.,  $P_{\rm D, u}(\Delta f) = P_{\rm D, d}(\Delta f)$ and $P_{\rm FA, u} = P_{\rm FA, d}$. In principle, the conjugate chirp preamble will be claimed to be detected if both the down- and up-chirp are detected. As a result, the preamble detection rate $P_{\rm D}$ can be computed as
\begin{equation}\label{P_D}
	P_{\rm D} = \int_{\Delta f_{\min}}^{\Delta f_{\max}} P_{\rm D,d}(x) P_{\rm D,u}(x) p(x) \, {\rm d}x, 
\end{equation} 
where $p(x)$ is the probability density function of the CFO. In our simulation experiments, we assume the CFO follows a uniform distribution between $\Delta f_{\min}$ and $\Delta f_{\max}$ and calculate its value numerically in MATLAB. On the other hand, the false alarm rate $P_{\rm FA}$ equals the product of two chirp signal false alarm rates, given by
\begin{equation} \label{P_FA}
	P_{\rm FA} =  \left(1 - \left(1-\exp\left(- \frac{\pi \gamma^2}{4}\right)\right)^{\rm W }\right)^2.
\end{equation}
With \eqref{P_FA}, the detection threshold $\gamma$ can be explicitly determined by a fixed false alarm rate $P_{\rm FA}$ and sliding window width $W$, expressed as \eqref{Eq-Prop-1}.
This completes the proof.

\section{Proof of Proposition~\ref{Proposition-2}} \label{Appendix-B}
 Since one entire DSSS-MSK signal correlation can be treated as the combination of chip-level MSK correlations, the correlation coefficient $\rho_{\rm DM}(\bm{d})$ can be rewritten as 
\begin{equation} \label{Eq-Appendix2-a}
	\rho_{\rm DM}(\bm{d}) = \frac{1}{K {\rm SF}_{\rm p}}\Big|\sum_{l=0}^{{\rm SF}_{\rm p} -1}\sum_{n = 0}^{K - 1}s_{\rm DM, 1}^{(l)}[n] \left( s_{\rm DM, 0}^{(l)}[n] \right)^{\ast}\Big|,
\end{equation}
where $s_{\rm DM,1}^{(l)}[n]$ is the $l^{\rm th}$ chip of DSSS-MSK symbol `1', expressed as 
\begin{equation}\label{i_MSK}
	s_{\rm DM,1}^{(l)}[n] =
	\left\{ \hspace{-5pt}
		\begin{array}{rl}	
	 \exp \left(j\frac{\pi d[l]}{2 K} n\right), & l = 0; \\
	\exp \left(j\frac{\pi}{2} \sum_{i=0}^{l-1}d[i]\right) 
	\exp \left(j\frac{\pi d[l]}{2K}n \right), & l > 0. \\
	\end{array} \right. \nonumber
\end{equation}
Therefore, \eqref{Eq-Appendix2-a} can be computed and given by
\begin{align} 
	\rho_{\rm DM}(\bm{d}) 
	&= \frac{1}{K {\rm SF}_{\rm p} \sin(\frac{\pi}{2K})}  \Big| \sum_{l=0}^{{\rm SF}_{\rm p} -1} \hspace{-5pt} (-1)^{l}\exp\left(j\pi\frac{K-1}{2K}d[l]\right) \Big|.  
\label{Eq-Appendix2-c}
\end{align}
If we want the phase rotation of each MSK chip to be canceled out by each other, this is equivalent to making $\rho_{\rm DM}(\bm{d})$ given by \eqref{Eq-Appendix2-c} equal zero. As a result, we must have 
\begin{equation}
	\sum_{l=0}^{{\rm SF}_{\rm p} -1} (-1)^{l}\exp\Big(j \pi \frac{K-1}{2K} d[l]\Big) = 0.
\label{Eq-Appendix2-d}
\end{equation}
Finally, $\eqref{Eq-Appendix2-d}$ can be readily reduced to the desired \eqref{DSSS_ort}.  
\end{appendices}

\bibliographystyle{IEEEtran}
\bibliography{reference}

\begin{thebibliography}{10}
\providecommand{\url}[1]{#1}
\csname url@samestyle\endcsname
\providecommand{\newblock}{\relax}
\providecommand{\bibinfo}[2]{#2}
\providecommand{\BIBentrySTDinterwordspacing}{\spaceskip=0pt\relax}
\providecommand{\BIBentryALTinterwordstretchfactor}{4}
\providecommand{\BIBentryALTinterwordspacing}{\spaceskip=\fontdimen2\font plus
\BIBentryALTinterwordstretchfactor\fontdimen3\font minus
  \fontdimen4\font\relax}
\providecommand{\BIBforeignlanguage}[2]{{%
\expandafter\ifx\csname l@#1\endcsname\relax
\typeout{** WARNING: IEEEtran.bst: No hyphenation pattern has been}%
\typeout{** loaded for the language `#1'. Using the pattern for}%
\typeout{** the default language instead.}%
\else
\language=\csname l@#1\endcsname
\fi
#2}}
\providecommand{\BIBdecl}{\relax}
\BIBdecl

\bibitem{mMTC}
Y.~Miao, W.~Li, D.~Tian, M.~S. Hossain, and M.~F. Alhamid, ``Narrowband
  {Internet of Things}: Simulation and modeling,'' \emph{IEEE Internet Things
  J.}, vol.~5, no.~4, pp. 2304--2314, Aug. 2017.

\bibitem{9778216}
R.~Marini, K.~Mikhaylov, G.~Pasolini, and C.~Buratti, ``Low-power wide-area
  networks: Comparison of {LoRaWAN} and {NB-IoT} performance,'' \emph{IEEE
  Internet Things J.}, vol.~9, no.~21, pp. 21\,051--21\,063, 2022.

\bibitem{11017626}
S.~Zhang, W.~Wen, P.~Wu, H.~Huang, L.~Zhu, Y.~Guo, T.~Yang, and M.~Xia,
  ``System-level simulation framework for {NB-IoT}: Key features and
  performance evaluation,'' \emph{IEEE Syst. J.}, vol.~19, no.~2, pp. 577--588,
  Jun. 2025.

\bibitem{10609524}
A.~Maleki, H.~H. Nguyen, E.~Bedeer, and R.~Barton, ``A tutorial on chirp spread
  spectrum modulation for {LoRaWAN}: Basics and key advances,'' \emph{IEEE Open
  J. Commun. Soc.}, vol.~5, pp. 4578--4612, 2024.

\bibitem{10683565}
A.~Maleki, E.~Bedeer, and R.~Barton, ``Performance evaluation and
  low-complexity detection of the {PHY} modulation of {LR-FHSS} transmission in
  {IoT} networks,'' in \emph{Proc. IEEE VTC'2024-Spring}, 2024, pp. 1--7.

\bibitem{9990567}
C.~Milarokostas, D.~Tsolkas, N.~Passas, and L.~Merakos, ``A comprehensive study
  on {LPWANs} with a focus on the potential of {LoRa/LoRaWAN} systems,''
  \emph{IEEE Commun. Surv. Tuts.}, vol.~25, no.~1, pp. 825--867, 2023.

\bibitem{1092333}
W.~Osborne and M.~Luntz, ``Coherent and noncoherent detection {CPFSK},''
  \emph{IEEE Trans. Commun.}, vol.~22, no.~8, pp. 1023--1036, Aug. 1974.

\bibitem{NPRACH}
Q.~Wu, P.~Wu, W.~Wen, T.~Yang, and M.~Xia, ``An efficient {NPRACH} receiver
  design for {NB-IoT} systems,'' \emph{IEEE Internet Things J.}, vol.~7,
  no.~10, pp. 10\,418--10\,426, Apr. 2020.

\bibitem{10844041}
X.~Cai, C.~Huang, P.~Chen, E.~Basar, and C.~Yuen, ``Design of non-coherent
  {RIS}-empowered {DCSK} with two-level nested index modulation,'' \emph{IEEE
  Trans. Wireless Commun.}, vol.~24, no.~4, pp. 3044--3058, Apr. 2025.

\bibitem{LPWAN2}
X.~Xiong, K.~Zheng, R.~Xu, W.~Xiang, and P.~Chatzimisios, ``Low power wide area
  machine-to-machine networks: key techniques and prototype,'' \emph{IEEE
  Commun. Mag.}, vol.~53, no.~9, pp. 64--71, 2015.

\bibitem{sx1272}
{Semtech Corp.}, ``{SX1272/73} - 860 {MHz} to 1020 {MHz} low power long range
  transceiver,'' 2019.

\bibitem{IEEE802154}
IEEE Std 802.15.4k, ``IEEE Standard for Low-Rate Wireless Networks'', 2020.

\bibitem{xu2017design}
R.~Xu, X.~Xiong, K.~Zheng, and X.~Wang, ``Design and prototyping of low-power
  wide area networks for critical infrastructure monitoring,'' \emph{IET
  Commun.}, vol.~11, no.~6, pp. 823--830, Dec. 2016.

\bibitem{DSSS_Diff}
S.~Shang, Y.~Hu, J.~Luo, and Y.~Wang, ``A new acquisition method based on
  differential correlation,'' in \emph{Proc. IEEE ICCSN'2016}, 2016, pp.
  196--200.

\bibitem{DC}
E.~{Hosseini} and E.~{Perrins}, ``Timing, carrier, and frame synchronization of
  burst-mode {CPM},'' \emph{IEEE Trans. Commun.}, vol.~61, no.~12, pp.
  5125--5138, Dec. 2013.

\bibitem{DC1}
D.~Park, C.~S. Park, and K.~Lee, ``Simple design of detector in the presence of
  frequency offset for {IEEE 802.15.4 LR-WPANs},'' \emph{IEEE Trans. Circuits
  Syst. II: Express Briefs}, vol.~56, no.~4, pp. 330--334, 2009.

\bibitem{MSK_Sync}
Q.~Lin, S.~Mohan, and M.~A. Weitnauer, ``Interference-insensitive
  synchronization scheme of {MSK} for transmit-only wireless sensor network,''
  in \emph{Proc. IEEE ICC'2016}, May 2016, pp. 1--6.

\bibitem{8657965}
Y.~Li, M.~Xia, and Y.-C. Wu, ``Activity detection for massive connectivity
  under frequency offsets via first-order algorithms,'' \emph{IEEE Trans.
  Wireless Commun.}, vol.~18, no.~3, pp. 1988--2002, Mar. 2019.

\bibitem{EPFL}
\BIBentryALTinterwordspacing
J.~Tapparel, ``Complete reverse engineering of {LoRa PHY},'' 2020. [Online].
  Available:
  \url{https://www.epfl.ch/labs/tcl/wp-content/uploads/2020/02/Reverse_Eng_Report.pdf}
\BIBentrySTDinterwordspacing

\bibitem{Seller2016}
O.~Seller and N.~Sornin., ``Low power long range transmitter,'' Feb. 2016, {US
  Patent 9.252,834 B2}.

\bibitem{9154273}
J.~Tapparel, O.~Afisiadis, P.~Mayoraz, A.~Balatsoukas-Stimming, and A.~Burg,
  ``An open-source {LoRa} physical layer prototype on {GNU Radio},'' in
  \emph{Proc. IEEE SPAWC'2020}, 2020, pp. 1--5.

\bibitem{9000820}
C.~Bernier, F.~Dehmas, and N.~Deparis, ``Low complexity {LoRa} frame
  synchronization for ultra-low power software-defined radios,'' \emph{IEEE
  Trans. Commun.}, vol.~68, no.~5, pp. 3140--3152, May 2020.

\bibitem{9501038}
M.~Xhonneux, O.~Afisiadis, D.~Bol, and J.~Louveaux, ``A low-complexity {LoRa}
  synchronization algorithm robust to sampling time offsets,'' \emph{IEEE
  Internet Things J.}, vol.~9, no.~5, pp. 3756--3769, 2022.

\bibitem{9148806}
O.~Afisiadis, A.~Burg, and A.~Balatsoukas-Stimming, ``Coded {LoRa} frame error
  rate analysis,'' in \emph{Proc. ICC'2020}, 2020, pp. 1--6.

\bibitem{Chirp_Sync1}
S.~Boumard and A.~Mammela, ``Time domain synchronization using {Newman} chirp
  training sequences in {AWGN} channels,'' in \emph{Proc. IEEE ICC'2005},
  vol.~2, 2005, pp. 1147--1151.

\bibitem{Chirp_Sync3}
A.~B. Martinez, A.~Kumar, M.~Chafii, and G.~Fettweis, ``A new approach for
  accurate time synchronization using chirp signals,'' in \emph{Proc. IEEE VTC
  Spring'2020}, May 2020, pp. 1--5.

\bibitem{4783007}
S.~Boumard and A.~Mammela, ``Robust and accurate frequency and timing
  synchronization using chirp signals,'' \emph{IEEE Trans. Broadcast.},
  vol.~55, no.~1, pp. 115--123, 2009.

\bibitem{6775034}
J.-C. Lin, ``Initial synchronization assisted by inherent diversity over
  time-varying frequency-selective fading channels,'' \emph{IEEE Trans.
  Wireless Commun.}, vol.~13, no.~5, pp. 2518--2529, 2014.

\bibitem{9083764}
A.~B. Martinez, A.~Kumar, M.~Chafii, and G.~Fettweis, ``A chirp-based frequency
  synchronization approach for flat fading channels,'' in \emph{Proc. 2nd 6G
  Wireless Summit}, 2020, pp. 1--5.

\bibitem{10498090}
G.~Xie, Q.~Liang, R.~Li, and D.~Li, ``Dual chirps direction shift keying
  modulation,'' \emph{IEEE Wireless Commun. Lett.}, vol.~13, no.~6, pp.
  1735--1739, 2024.

\bibitem{9828505}
A.~W. Azim, A.~Bazzi, R.~Shubair, and M.~Chafii, ``Dual-mode chirp spread
  spectrum modulation,'' \emph{IEEE Wireless Commun. Lett.}, vol.~11, no.~9,
  pp. 1995--1999, Sep. 2022.

\bibitem{10183362}
A.~W. Azim, A.~Bazzi, M.~Fatima, R.~Shubair, and M.~Chafii, ``Dual-mode time
  domain multiplexed chirp spread spectrum,'' \emph{IEEE Trans. Veh. Technol.},
  vol.~72, no.~12, pp. 16\,086--16\,097, Dec. 2023.

\bibitem{10296020}
X.~Yu, X.~Cai, W.~Xu, H.~Sun, and L.~Wang, ``Differential phase shift
  keying-aided multi-mode chirp spread spectrum modulation,'' \emph{IEEE
  Wireless Commun. Lett.}, vol.~13, no.~2, pp. 298--302, Feb. 2024.

\bibitem{10391276}
A.~W. Azim, R.~Shubair, and M.~Chafii, ``Chirp spread spectrum-based waveform
  design and detection mechanisms for {LPWAN}-based {IoT}: A survey,''
  \emph{IEEE Access}, vol.~12, pp. 24\,949--25\,017, 2024.

\bibitem{FSK_DSSS}
M.-K. Oh and C.-h. Shin, ``{FSK} {PHY} design with enhanced synchronization for
  low-power wide-area connectivity,'' in \emph{Proc. IEEE ICTC'2017}, 2017, pp.
  1257--1259.

\bibitem{8723130}
M.~Chiani and A.~Elzanaty, ``On the {LoRa} modulation for {IoT}: Waveform
  properties and spectral analysis,'' \emph{IEEE Internet Things J.}, vol.~6,
  no.~5, pp. 8463--8470, Oct. 2019.

\bibitem{9555814}
T.~Ameloot, H.~Rogier, M.~Moeneclaey, and P.~Van~Torre, ``{LoRa} signal
  synchronization and detection at extremely low signal-to-noise ratios,''
  \emph{IEEE Internet Things J.}, vol.~9, no.~11, pp. 8869--8882, Jun. 2022.

\bibitem{CFAR}
L.~L. Scharf and C.~Demeure, \emph{Statistical Signal Processing: Detection,
  Estimation, and Time Series Analysis}.\hskip 1em plus 0.5em minus 0.4em\relax
  Prentice Hall, 1991.

\bibitem{8903531}
O.~Afisiadis, M.~Cotting, A.~Burg, and A.~Balatsoukas-Stimming, ``On the error
  rate of the {LoRa} modulation with interference,'' \emph{IEEE Trans. Wireless
  Commun.}, vol.~19, no.~2, pp. 1292--1304, 2020.

\bibitem{Estimation}
S.~M. Kay, \emph{Fundamentals of Statistical Signal Processing: Estimation
  Theory}.\hskip 1em plus 0.5em minus 0.4em\relax Prentice-Hall, Inc., 1993.

\bibitem{semtech2015an1200}
{Semtech Corp.}, ``{AN1200.22 LoRa} modulation basics,'' 2015.

\bibitem{FSK_Enhanced}
Y.~Xiang, R.~Okumura, K.~Mizutani, and H.~Harada, ``Data rate enhancement of
  {FSK} transmission scheme for {IEEE} 802.15.4-based field area network,''
  \emph{IEEE Sensors J.}, vol.~21, no.~7, pp. 9600--9611, 2021.

\bibitem{X310}
\BIBentryALTinterwordspacing
``{Ettus Research X300/X310}.'' [Online]. Available:
  \url{http://kb.ettus.com/X300/X310}
\BIBentrySTDinterwordspacing

\bibitem{36101}
3GPP, TS 36.101: ``User Equipment (UE) radio transmission and reception", 2018.

\bibitem{Detection}
S.~M. Kay, \emph{Fundamentals of Statistical Signal Processing: Detection
  Theory}.\hskip 1em plus 0.5em minus 0.4em\relax Prentice-Hall, Inc., 1993.

\end{thebibliography}

\begin{IEEEbiography}
	[{\includegraphics[width=1in, height=1.25in, clip, keepaspectratio]{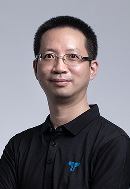}}]{Wenkun Wen} (Member, IEEE) received the Ph.D. degree in Telecommunications and Information Systems from Sun Yat-sen University, Guangzhou, China, in 2007. Since 2020, he has been with Techphant Technologies Co. Ltd., Guangzhou, China, as Chief Engineer.

From 2008 to 2009, he was with the Guangdong-Nortel R\&D center in Guangzhou, China, where he worked as a system engineer for 4G systems. From 2009 to 2012, he worked at the LTE R\&D center of New Postcom Equipment Co. Ltd., Guangzhou, China, where he served as the 4G standard team manager. From 2012 to 2018, he was with the 7th Institute of China Electronic Technology Corporation (CETC) as an expert in wireless communications. From 2018 to 2020, he served as Deputy Director of the 5G Innovation Center at CETC. His research interests include 5G/B5G mobile communications, machine-type communications, narrow-band wireless communications, and signal processing.
\end{IEEEbiography}	

\vfill

\begin{IEEEbiography}
	[{\includegraphics[width=1in, height=1.25in, clip, keepaspectratio]{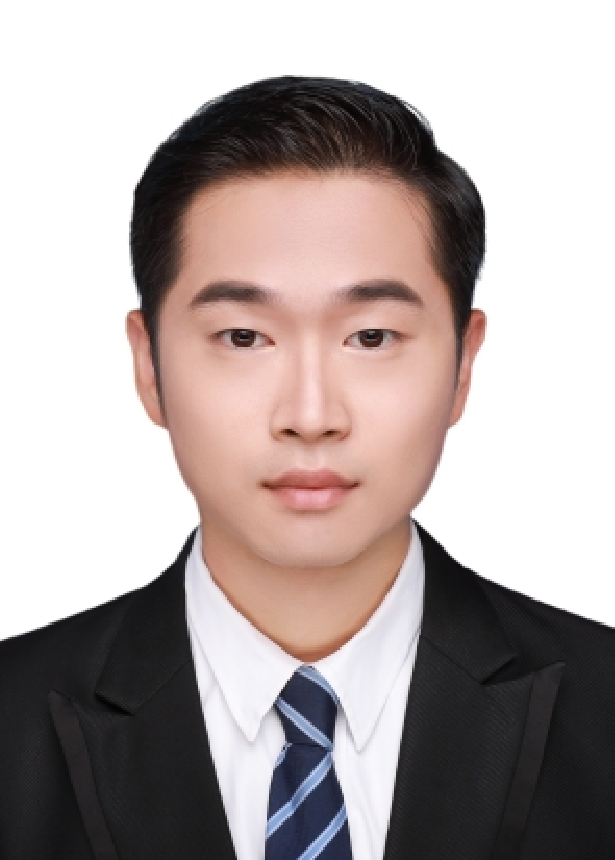}}]{Ruiqi Zhang} received the B.S. degree in communication engineering and the M.S. degree in information and communication engineering from Sun Yat-sen University, Guangzhou, China, in 2019 and 2022, respectively. Since 2022, he has been with the China Telecom Research Institute, Guangzhou, China. His research interests include 5G/6G mobile communications and the Internet of Things.
\end{IEEEbiography}

\vfill

\begin{IEEEbiography}
	[{\includegraphics[width=1in, height=1.25in, clip, keepaspectratio]{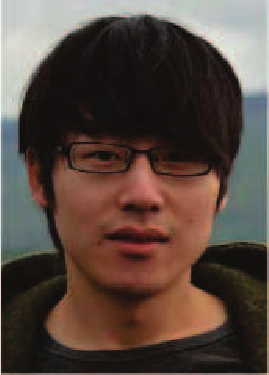}}]{Peiran Wu} (Member, IEEE) received the Ph.D. degree in electrical and computer engineering from The University of British Columbia (UBC), Vancouver, Canada, in 2015.
	
	From October 2015 to December 2016, he was a Post-Doctoral Fellow at UBC. In the Summer of 2014, he was a Visiting Scholar with the Institute for Digital Communications, Friedrich-Alexander-University Erlangen-Nuremberg (FAU), Erlangen, Germany. Since February 2017, he has been with Sun Yat-sen University, Guangzhou, China, where he is currently an Associate Professor. Since 2019, he has been an Adjunct Associate Professor with the Southern Marine Science and Engineering Guangdong Laboratory, Zhuhai, China. His research interests include mobile edge computing, wireless power transfer, and energy-efficient wireless communications. 
\end{IEEEbiography}

\begin{IEEEbiography}
	[{\includegraphics[width=1in, height=1.25in, clip, keepaspectratio]{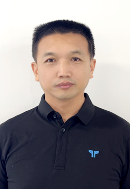}}]{Tierui Min } received the M.S. degree in Computer Application Technology from Sichuan University, Chengdu, China, in 2004. Since 2021, he has been with Techphant Technologies Co. Ltd., Guangzhou, China, as a system engineer.

From 2004 to 2010, he was with the Guangdong-Nortel R\&D center in Guangzhou, China, where he worked as a system engineer for 4G systems. From 2010 to 2012, he worked at the LTE R\&D center of New Postcom Equipment Co. Ltd., Guangzhou, China,  where he served as a system engineer. From 2012 to 2018, he was with the 7th Institute of China Electronic Technology Corporation (CETC) as a system engineer in wireless communications. From 2018 to 2021, he served as a system engineer at the 5G Innovation Center at CETC. His research interests include 5G/B5G mobile communications and narrow-band wireless communications.
\end{IEEEbiography}

\vfill
	
\begin{IEEEbiography}
	[{\includegraphics[width=1in, height=1.25in, clip, keepaspectratio]{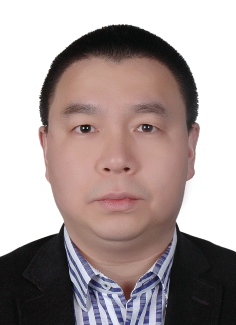}}]{Minghua Xia} (Senior Member, IEEE) received the Ph.D. degree in Telecommunications and Information Systems from Sun Yat-sen University, Guangzhou, China, in 2007.
	
	From 2007 to 2009, he was with the Electronics and Telecommunications Research Institute (ETRI) of South Korea, Beijing R\&D Center, Beijing, China, where he worked as a member and then as a senior member of the engineering staff. From 2010 to 2014, he was in sequence with The University of Hong Kong, Hong Kong, China; King Abdullah University of Science and Technology, Jeddah, Saudi Arabia; and the Institut National de la Recherche Scientifique (INRS), University of Quebec, Montreal, Canada, as a Postdoctoral Fellow. Since 2015, he has been a Professor at Sun Yat-sen University. Since 2019, he has also been an Adjunct Professor with the Southern Marine Science and Engineering Guangdong Laboratory (Zhuhai). His research interests are in the general areas of wireless communications and signal processing.
\end{IEEEbiography}

\end{document}